\documentclass[%
      aps,
      prb, 
      twocolumn,
      showkeys,
      superscriptaddress,
      showpacs]{revtex4-2}
      
\usepackage[utf8]{inputenc}
\usepackage[T1]{fontenc}
\usepackage{graphicx}
\graphicspath{ {figures/} }
\usepackage{amsmath, amsfonts, amssymb, mathtools}	

\begin{document}


\title{Designable exciton mixing through layer alignment in WS\textsubscript{2}-graphene heterostructures}

\author{Amir Kleiner}

\affiliation{Department of Molecular Chemistry and Materials Science, Weizmann Institute of Science, Rehovot 7610001, Israel}

\author{Daniel \surname{Hernang\'{o}mez-P\'{e}rez}}

\affiliation{Department of Molecular Chemistry and Materials Science, Weizmann Institute of Science, Rehovot 7610001, Israel}

\author{Sivan Refaely-Abramson}

\affiliation{Department of Molecular Chemistry and Materials Science, Weizmann Institute of Science, Rehovot 7610001, Israel}

\email[Corresponding author:]{sivan.refaely-abramson@weizmann.ac.il}



\date{\today}

\begin{abstract}
    Optical properties of heterostructures composed of layered 2D materials, such as transition metal dichalcogenides (TMDs) and graphene, are broadly explored. Of particular interest are light-induced energy transfer mechanisms in these materials and their structural roots.
    Here, we use state-of-the-art first-principles calculations to study the excitonic composition and the absorption properties of WS\textsubscript{2}--graphene heterostructures as a function of interlayer alignment and the local strain resulting from it. We find that Brillouin zone mismatch and the associated energy level alignment between the graphene Dirac cone and the TMD bands dictate an interplay between interlayer and intralayer excitons, mixing together in the many-body representation upon the strain-induced symmetry breaking in the interacting layers. Examining the representative cases of the  0$^\circ$ and 30$^\circ$ interlayer twist angles, we find that this exciton mixing strongly varies as a function of the relative alignment. We quantify the effect of these structural modifications on exciton charge separation between the layers and the associated graphene-induced homogeneous broadening of the absorption resonances.
    Our findings provide guidelines for controllable optical excitations upon interface design and shed light on the importance of many-body effects in the understanding of optical phenomena in complex heterostructures.
\end{abstract}

\maketitle

\section{Introduction}
Two-dimensional layered van der Waals heterostructures have been extensively studied for a wealth of applications in photophysics. Their structural complexity provides adjustable and designable light-matter features, offering the generation of controllable optical states for application in optoelectronics, light harvesting, and material-based information science.
~\cite{geim_van_2013, novoselov_2d_2016, liu_van_2016, massicotte_picosecond_2016, jin_ultrafast_2018, kennes_moire_2021, lemme_2d_2022}. 
Of particular interest are layered heterostructures (HSs) composed of graphene (Gr) and transition metal dichalcogenides (TMDs)~\cite{jin_tuning_2015, hill_exciton_2017, aeschlimann_direct_2020, krause_microscopic_2021}. 
These interfaces combine the semi-metallic nature, large charge carrier mobility, and constant infrared optical absorption of Gr ~\cite{novoselov_electric_2004, castro_neto_electronic_2009, yang_excitonic_2009,  mak_measurement_2008, das_sarma_electronic_2011,  mak2014tuning} with the semiconducting nature of TMDs, characterized by strong exciton binding and significant spin-orbit coupling~\cite{mak_atomically_2010,  splendiani_emerging_2010,ramasubramaniam_large_2012, ugeda_giant_2014, chernikov_exciton_2014, hanbicki_measurement_2015}. 
Reflectance contrast measurements in these interfaces show line broadening~\cite{hill_exciton_2017} that can be assigned to charge transfer processes between the coupled layers~\cite{song_ultrafast_2018, krause_ultrafast_2021,  trovatello_ultrafast_2022, fu_long-lived_2021}. Recent time-resolved angle-resolved photoemission spectroscopy (tr-ARPES) studies further suggest a relation between interlayer coupling and reciprocal-space spreading of the optical transitions~\cite{dong_observation_2023}. 
These processes are particularly affected by intrinsic atomic reconstruction upon interlayer twisting~\cite{weston_atomic_2020, rosenberger_twist_2020}.   

From a theoretical point of view, Gr-TMD HSs serve as a fascinating test case to establish a direct relation between the intrinsic structural disorder induced by layer stacking, which is sensitive to interlayer alignment, and the observed optical signature and spectral broadening. 
The presence of the Gr Dirac cone within the TMD band gap~\cite{pierucci_band_2016, wilson_determination_2017, henck_electronic_2018} opens pathways to the tunability of the electronic structure upon interface alignment and layer thickness, well known to modify the magnitude and directness of the TMD band gap~\cite{scalise_strain-induced_2012, yun_thickness_2012, ebnonnasir_tunable_2014, junior_proximity-enhanced_2023}. 
The effect of Gr on the TMD electronic levels is typically associated with non-local dielectric screening, manifested in band gap renormalization~\cite{magnozzi_optical_2020, riis-jensen_electrically_2020, tebbe_tailoring_2023} and subsequent reduction in the excitonic binding energies~\cite{latini_excitons_2015}. However, a first-principles understanding of the change in the exciton states themselves upon layer interfacing is yet to be achieved. 

Recent studies used many-body perturbation theory to show that upon TMD-TMD heterostructure composition, excitons include a wealth of electron-hole transitions with varying spin and momentum properties~\cite{karni2022structure, naik_intralayer_2022, barre_optical_2022}, leading to the emergence of hybridized excitons with mixed interlayer and intralayer nature. 
Such state mixing results from local symmetry breaking and resonating excitation energies,  manifesting many-body effects associated with the relative reciprocal space alignment of the layers, as well as the interface-induced atomic reconstruction and the associated local strain ~\cite{kundu2023exciton, kundu2022moire}. 
Recently we have explored the effect of atomic defects on the optical properties in Gr-TMD HSs~\cite{hernangomez-perez_reduced_2023}, finding that largely mixed subgap, defect-induced excitons dominate the absorption, a property that in the monolayer TMD case leads to greatly reduced valley selectivity~\cite{refaely-abramson_defect-induced_2018, amit_tunable_2022}.
The optical activity of Gr-TMD junctions is thus closely coupled to their interfacial structural details through the many-body exciton nature.
A first-principles understanding of the exciton state mixing as a function of interface layer alignment in Gr-TMD HSs is then of great interest and can shed light on the underlying mechanisms dominating the excited-state phenomena in these systems.

In this work, we study the effect of layer alignment on optical and excitonic properties in WS\textsubscript{2}--graphene heterobilayers using many-body perturbation theory.
Exploring two representative configurations of this interface, with both 0$^\circ$ and 30$^\circ$ interlayer twist angles, we show that optical excitations largely mix electron-hole transitions within the TMD and Gr layers and between them. This mixing depends on the interfacial configuration through the angle-dependent band alignment. As a result, states that are optically active due to their TMD layer contribution also have significant contributions from the Gr layer. 
We analyze the exciton charge separation associated with this effect and investigate the absorption line broadening resulting from the large amount of excitons contributing to the optical resonances.
Our findings suggest that charge transfer mechanisms in these heterostructures are largely tunable with the twist angle, with significant charge separation occurring already upon light absorption and setting a coherent starting point to charge transfer dynamics, e.g., by efficient scattering into graphene-induced low-lying states.  

\section{Results and Discussion}
\begin{figure}[t]
    \centering
    \includegraphics[width=1.0\linewidth]{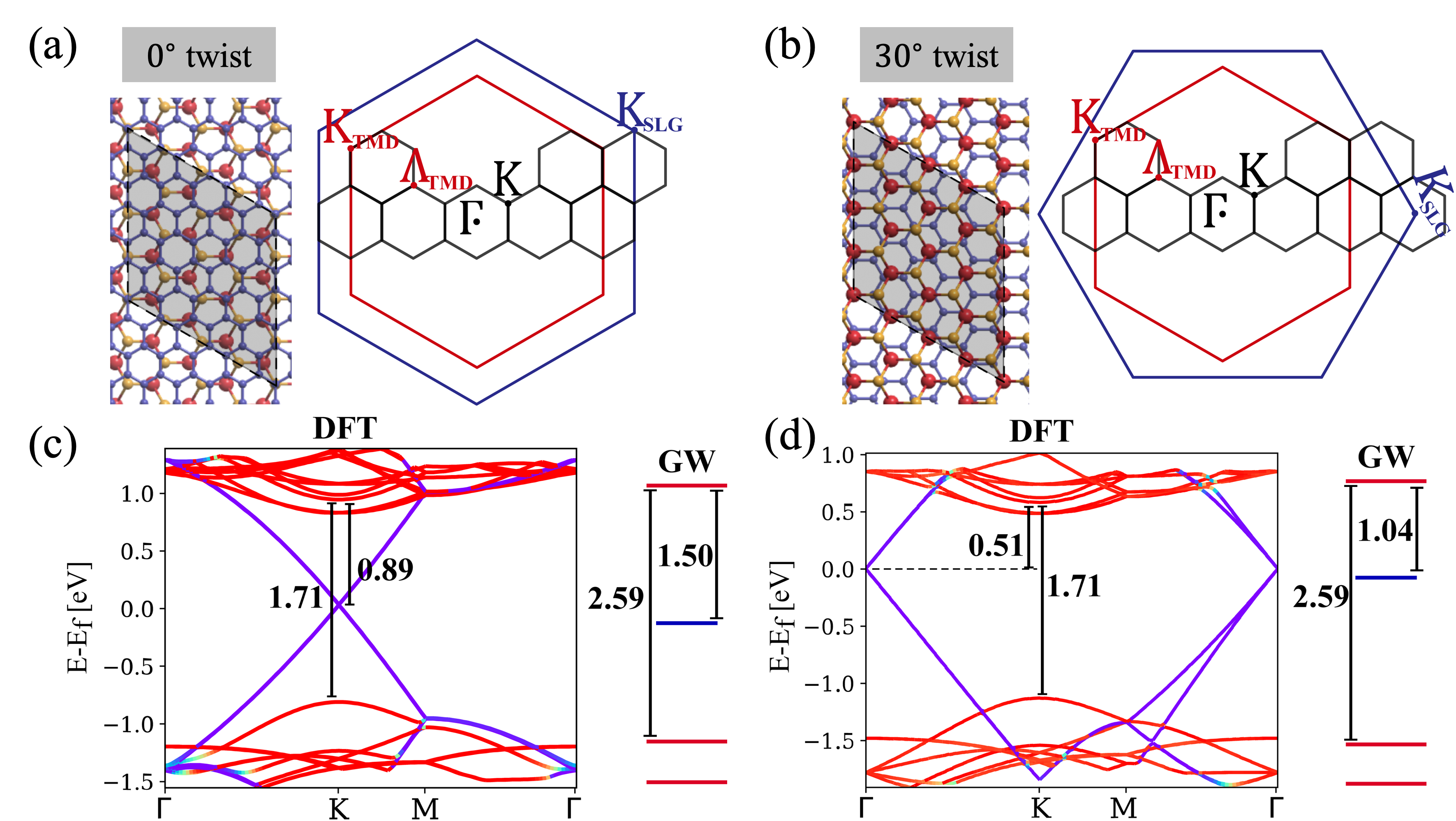}
    \caption{(a,b) Atomic structure of the studied WS\textsubscript{2}-Gr HS with $0^\circ$ and $30^\circ$ interlayer twist angles (left);  periodic supercell is shown in grey, with the corresponding BZs (right) of WS\textsubscript{2} (red) and graphene (blue) and the mini-BZ of the supercell (black).
    (c,d) DFT band structures of the $0^\circ$ and $30^\circ$ twisted HS (left) and the computed GW energy levels (right) of the TMD gap and its alignment with the graphene Dirac point. The TMD band gap and the TMD-graphene energy gap are shown in eV. The color code represents the layer contribution to each band, ranging from pure TMD (red) to pure graphene (purple).}
    \label{fig:struct}
\end{figure}
 
The studied heterostructures are presented in Fig.~\ref{fig:struct}(a,b).
We use commensurate periodic supercells with $4\times4$ repeating elementary units of the TMD layer and $5\times5$ repeating elementary units of the Gr layer for the $0^\circ$-twist angle; and $3\sqrt{3}\times3\sqrt{3}$ repeating elementary units of the Gr Layer for the $30^\circ$-twist angle (Fig.~\ref{fig:struct}(a,b), left). These alignments result in a minimal strain of $2.6\%$ for the $0^\circ$-twist angle and $1.3\%$ for the $30^\circ$-twist angle in the Gr layer, respectively, and no strain in the TMD layer. However, local strain is induced in the TMD upon structural relaxation as a function of the Gr-TMD layer alignment (see SI for full details). 
The unit cell Brillouin zones (BZ), corresponding to the separate layers, fold in this configuration onto the smaller, shared mini-BZ (mBZ) of the supercell (Fig.~\ref{fig:struct}(a,b), right).
The TMD bands fold identically in both configurations (due to the equivalent TMD-layer BZ in both cases), resulting in the TMD minimal gap at the $K$  point (or equivalently $K'$; we consider only $K$ in this work for brevity). The conduction-band minimum is at the $\Lambda$ point of the unit cell (located midway between $K$ and $\Gamma$),  folded onto the $K$ point of the mBZ.
The different twist configurations lead to different folding of the graphene bands in the mBZ of the two HSs: while in the $0^\circ$ HS the Dirac point folds to the mBZ $K$ point, in the $30^\circ$ HS, it folds to the mBZ $\Gamma$ point.
As we show below, this enforces substantial differences in the allowed optical transitions for both systems.

We use density functional theory (DFT)\cite{kohn_self-consistent_1965} to calculate the electronic band structure of the two systems, as shown in the left panels of Fig.~\ref{fig:struct}(c,d) along the $\Gamma$-$K$-$M$-$\Gamma$ path of the mBZ. Band colors give the relative contribution of the graphene layer (purple) and the TMD layer (red), with some states hybridized at band crossings (see SI for further details). 
In completion with the BZ folding scheme, in the case of the $0^\circ$-twist HS (Fig.~\ref{fig:struct}(c)), the graphene Dirac cone is aligned in reciprocal space with the TMD direct band gap, both folded onto the K point of the mBZ. For the $30^\circ$-twist HS (Fig.~\ref{fig:struct}(d)), the Dirac cone is centered at the $\Gamma$ point, and the minimal gap of the TMD occurs at the $K$ point of the mBZ. Importantly, the DFT electronic energy-level alignment of the Dirac point within the gap is different in the two systems, located $0.89$~eV and $0.51$~eV below the TMD conduction band minimum at $K$ for $0^\circ$ and $30^\circ$-twisted HSs, respectively.

\begin{figure*}
    \centering
    \includegraphics[width=1.0\linewidth]{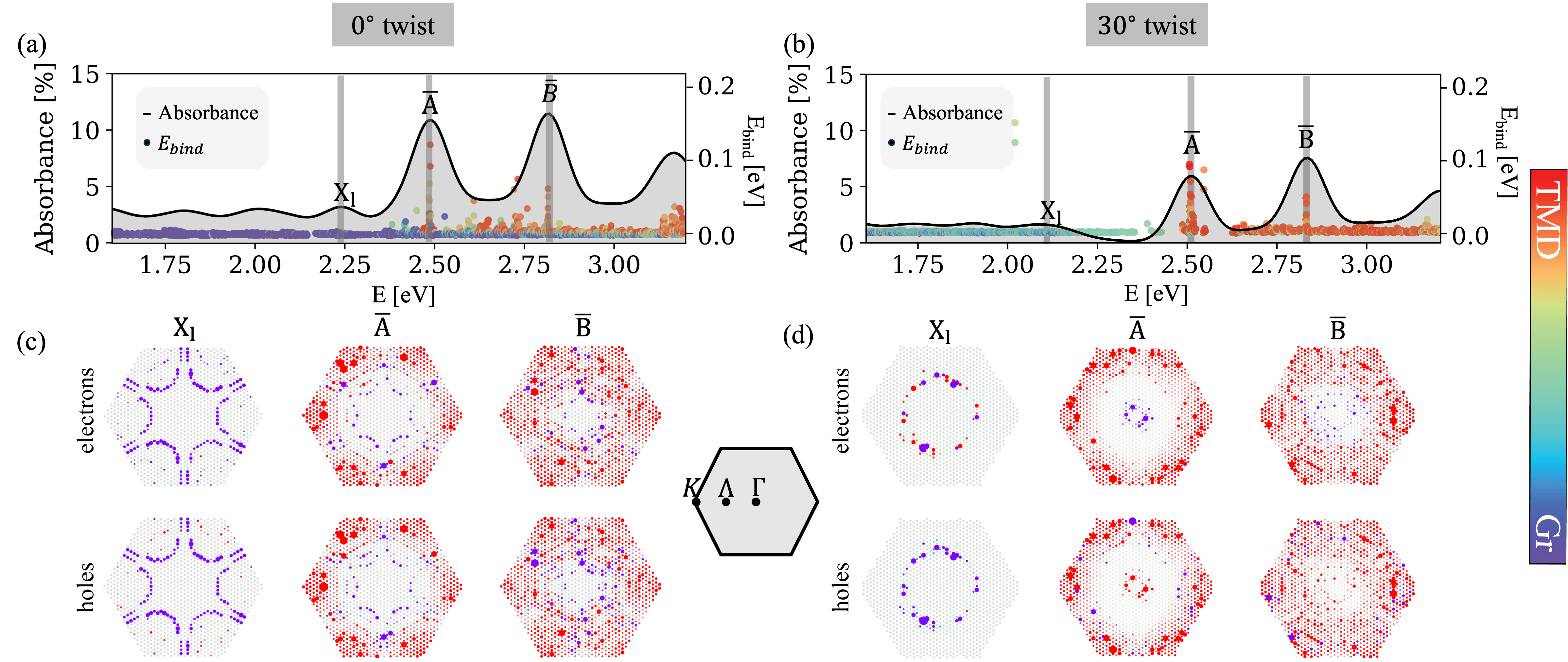}
    \caption{
    (a,b) Absorbance spectra of the $0^\circ$ and $30^\circ$ twisted HSs (black line). The binding energies of each optically active exciton are shown with dots, with colors representing the layer contribution of each excitation, from pure TMD (red) to pure graphene (purple).
    (c,d) Momentum-space representation of selected excitons, corresponding to the shaded vertical grey lines in panels (a,b).  Lower and upper panels show the hole and electron distribution in the mBZ, respectively.
    }
    \label{fig:optics}
\end{figure*}

We further apply many-body perturbation theory within the G$_0$W$_0$ approximation~\cite{hybertsen_electron_1986}, using the DFT electronic states and energies as a starting point, to obtain corrected quasiparticle energy levels as shown in the right panels of Fig.~\ref{fig:struct}(c,d). The resulting TMD GW band gaps are $2.59$~eV in both systems. 
Our computed GW gaps are $\sim 10-20\%$ larger than the experimental band gaps reported for WS\textsubscript{2}--Gr heterobilayers, of $2.1-2.3$~eV \cite{raja_coulomb_2017,waldecker_rigid_2019, trovatello_ultrafast_2022, song_ultrafast_2018, yuan_photocarrier_2018, giusca_probing_2019, krause_ultrafast_2021, aeschlimann_direct_2020, tebbe_tailoring_2023}. While still within a reasonable error bar, this difference is larger than expected. We assign it to both further surface renormalization in experiment, as well as to a computational effect of state hybridization in the mBZ, as was recently observed in other systems as well~\cite{kundu2023exciton}.
The difference in the Dirac cone alignment within the TMD gap remains, now located $1.50$~eV and $1.04$~eV below the TMD conduction band minimum at K, for $0^\circ$ and $30^\circ$-twisted HSs, respectively. We note recent experimental observations supporting the Gr Fermi level being closer to the conduction region~\cite{song_ultrafast_2018,krause_microscopic_2021}.
To evaluate the graphene-induced change in the band gap, we compare our results with a similar GW calculation of an isolated TMD layer that includes the atomic reconstruction due to graphene. This results in a gap renormalization of $\sim$0.3 eV in both systems (see SI), in good agreement with experiment~\cite{song_ultrafast_2018, trovatello_ultrafast_2022, raja_coulomb_2017, coy_diaz_direct_2015, krause_ultrafast_2021}. 

Next, we explore the optical spectra and excitonic states of the examined HSs by employing the Bethe-Salpeter equation (BSE)~\cite{rohlfing_electron-hole_2000, hybertsen_electron_1986} within many-body perturbation theory. 
Fig.~\ref{fig:optics}(a) shows the calculated absorbance of the $0^\circ$ HS (black curve).  
On top of the absorbance, we show the binding energies of all the excitons in this energy range with finite oscillator strength ($\mu_X$, here chosen as $\mu_X$>$1$). The binding energy is computed by subtracting the energy of each exciton ($\Omega_X$) from the quasiparticle energy gaps $E_{c\mathbf{k}} - E_{v\mathbf{k}}$, considering all the electron ($v$) - hole ($c$) transitions composing the exciton state at each \textbf{k}-point, weighted by the exciton coefficients obtained from the solution of the BSE ($A^X_{vc\mathbf{k}}$) and summed over all the BZ, through 
$ E^X_\textnormal{bind} =\sum_{vc\mathbf{k}}|A^X_{vc\mathbf{k}}|^2 (E_{c\mathbf{k}} - E_{v\mathbf{k}}) - \Omega_X$ (see SI for full details).
Dot colors represent the relative contribution of each layer to each exciton: intralayer excitons located solely at the graphene layer (purple), intralayer excitons located solely at the TMD layer (red), and excitons delocalized between both layers (ranging in between).
For the $0^\circ$ twisted HS, the main excitons composing the absorbance peaks have binding energies $\sim$100 meV, reduced in comparison to the case of pristine TMDs, as was also observed before~\cite{latini_excitons_2015, trovatello_ultrafast_2022, rosner_two-dimensional_2016, tebbe_tailoring_2023}.
Transitions at lower energies oscillate around the constant absorbance of graphene at $2.4\%$~\cite{stauber_optical_2008,nair_fine_2008,mak_measurement_2008,yang_excitonic_2009,magnozzi_optical_2020, hernangomez-perez_reduced_2023}, and are primarily composed of intralayer graphene states. 
The two pronounced peaks at $2.49$ eV and at $2.82$ eV, referred as $\bar{A}$ and $\bar{B}$, are mainly intralayer TMD excitons, with a slight mixing of graphene states. 
These states are associated with the well-known $A$ and $B$ peaks of TMDs, here also mixing additional graphene-containing transitions, as well as transitions corresponding to the excitation of the TMD $\Lambda$ point, in agreement with recent experimental observations \cite{dong_observation_2023}.

Interestingly, in the case of the $30^\circ$-twist HS (Fig.~\ref{fig:optics}(b), our results show a very different absorbance composition. 
The low-energy states oscillate below the graphene constant absorption value in the infrared optical range and consequently have significant interlayer contributions. 
We observe the main TMD-like $\bar{A}, \bar{B}$ peak features at similar energies of $2.50$ eV and $2.83$ eV. 
The substantially reduced absorbance between the low-energy region and the $\Tilde{\textnormal{A}}$ peak points to the fully interlayer-nature of the excitons in this region, in which intralayer graphene contributions are no longer present while intralayer TMD contributions are not yet present, thus decreasing the oscillator strength of these low-lying states. 
In both systems, we also observe dark excitons with relatively large binding energies, but with no signature in absorbance. Additionally, some of the strongly bound excitons exhibit a substantial contribution from the graphene layer.

To further analyze the excitations composing the complex absorbance picture described above, we plot the \textbf{k}-resolved exciton compositions in the energy regions of selected absorbance peaks, Fig.~\ref{fig:optics}(c,d).
In each energy region, the bright exciton states are shown in an interval of $\pm 5$ meV around the maximum of the absorption peaks. We show the integrated contribution of the holes (lower panels) and electrons (upper panels) separately, weighted by the exciton oscillator strength (see SI for further analysis). 
This representation allows further insights into the different features of the two structures. Notably, the \textbf{k}-space spreading of the low-energy exciton ($X_l$) greatly differs between the two examined structures. For the $0^\circ$ twist angle, the involved electron-hole transitions are primarily of intralayer graphene nature. The \textbf{k}-space spreading around the $K$, $K'$ valleys is a result of the low-lying excitation near the Dirac point in this structure.  In contrast, for the $30^\circ$ twist angle structure, the low-lying excitation has an interlayer graphene-TMD nature. It is also centered around the Dirac cone, this time at the $\Gamma$ region. This is a direct manifestation of the electronic band-alignment effect on the optical transitions. 
We note in passing that though the observed optical transitions are momentum-direct in the mBZ, some are not expected in the unfolded BZ (UBZ) framework. This property depends on the BZ folding associated with the specific twist angle and was recently shown by some of us to induce unexpected optical transitions also in twisted TMD-TMD heterobilayers ~\cite{kundu2023exciton}.
The $\bar{\textnormal{A}}$ peak region in both systems is mainly composed of intralayer TMD transitions around the $K$, $K'$ points of the mBZ; in both states, further mixing occurs with intralayer graphene and interlayer TMD-graphene transitions in the $0^\circ$ and $30^\circ$ twist structures, respectively.  The $\bar{\textnormal{B}}$ peak region shows large mixing of graphene-graphene, TMD-TMD, graphene-TMD, and TMD-graphene transitions.

The above findings point to the strongly mixed interlayer and intralayer nature of the excitations, which strongly depends on the band alignment resulting from the twist angle.   
We note that while the mBZ is determined by the choice of the periodic supercell for each structure, we consider the examined structures as good test cases to emphasize these delicate structural dependencies, even though experimental results may divert from our specific predictions upon structural variations.  
The binding energies of the various excitons manifest their intralayer TMD contribution; interlayer TMD-Gr and intralayer Gr-Gr transitions have smaller binding energies. Importantly, strongly mixed states induce bound excitons that hold some interlayer nature - of great interest for charge transfer decay processes proceeding the light excitation. 
In the following, we further analyze the twist-induced excitonic nature and, in particular, its effect on exciton charge separation and spectral broadening.

\begin{figure}
        \centering
        \includegraphics[width=1.0\linewidth]{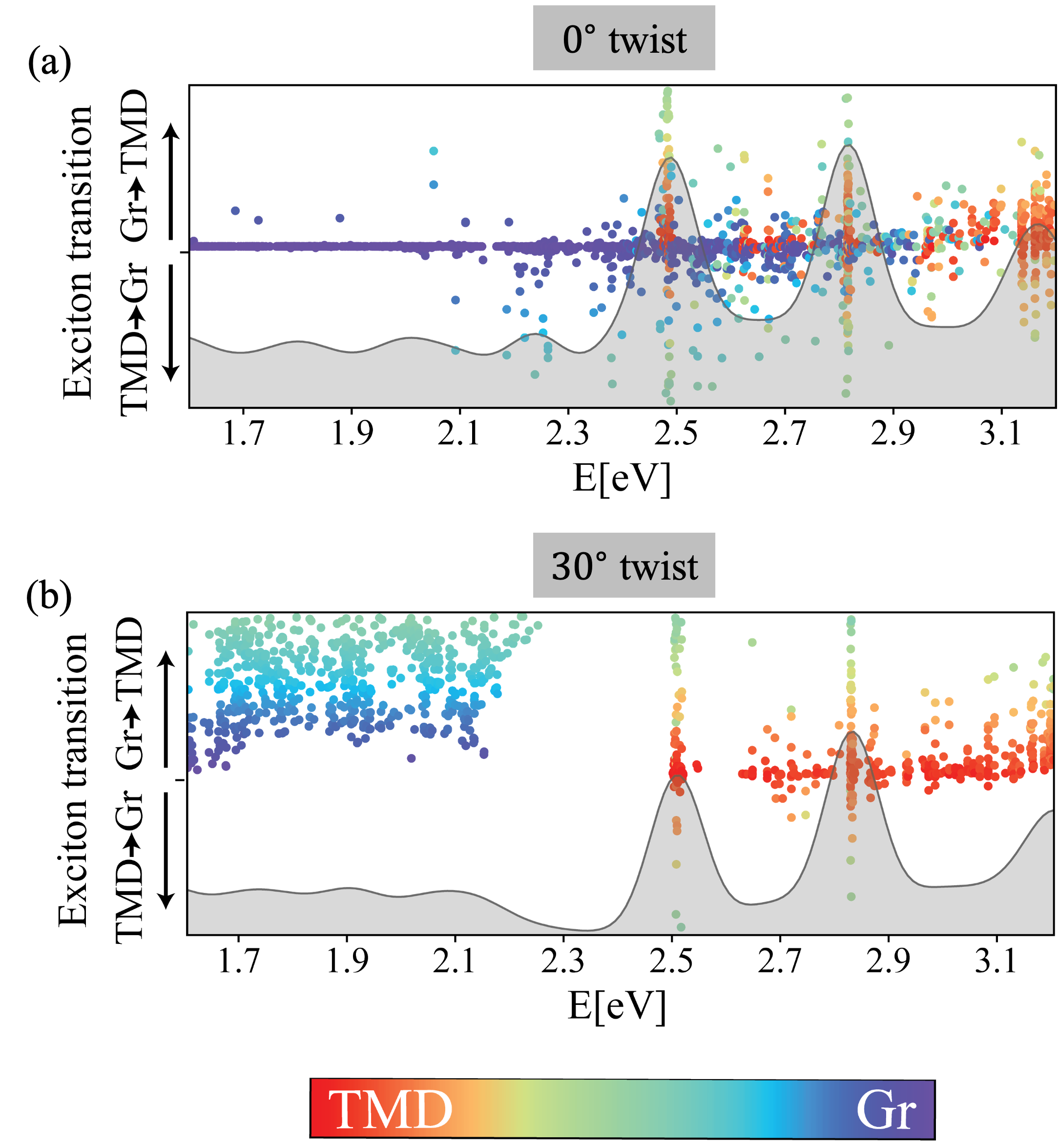}
        \caption{Measure of the interlayer charge separation for each exciton state in the examined (a) $0^\circ$ and (b) $30^\circ$ twisted HSs. Negative values are for excitons with holes located at the TMD layer and electrons at the graphene layer, and positive values for opposite transitions.
        Dot colors represent the overall layer contribution of each exciton state, with purple corresponding to excitons located at the graphene layer only, red to excitons located at the TMD layer only, and excitons delocalized in both layers spread in between.
        Only excitons with oscillator strength  $\mu_X$>$5$ are shown (see SI for lower thresholds).
        }
        \label{fig:natures}
    \end{figure}

For a numerical determination of the exciton charge-separation nature, Fig.~\ref{fig:natures} shows an analysis of the interlayer electron-hole contribution per exciton in both structures.
We differentiate between excitons with electrons and holes localized in the same layer - namely intralayer excitons, and those with electrons and holes localized in different layers - namely interlayer excitons (see SI).
We specify the directionality of these transitions, with a distinction between excitons with holes located on the Gr and electrons on the TMD layer (Gr $\rightarrow$ TMD) and those with holes located on the TMD and electrons on the Gr layer (TMD $\rightarrow$ Gr).
Each dot corresponds to an exciton state, with colors representing its overall layer contribution. The absorption spectra are presented in the background.
Our analysis shows a clear distinction in the interlayer exciton nature between the two structures. For the $0^\circ$ twist HS, 
Fig.~\ref{fig:natures}(a), most of the low-lying excitations are of intralayer graphene nature. The exciton charge separation mostly occurs around the $\bar{A}$ and $\bar{B}$ peaks and spreads rather evenly in both directions.
A very different picture emerges, however, for the $30^\circ$ twisted HS, Fig.~\ref{fig:natures} (b), in which the charge separation - occurring already in the low-lying excitation energies - has preferred directionality, with holes located in the graphene layer and electrons in the TMD layer. This property remains in the $\bar{A}$ and $\bar{B}$ peak regions. In this case, intralayer graphene excitons barely participate in the absorption. These results emphasize how a small change in the interlayer alignment can lead to substantial effects on the excitonic nature - and in particular, on the charge separation occurring already upon light absorption.

\begin{figure}
    \centering
    \includegraphics[width=0.9\linewidth]{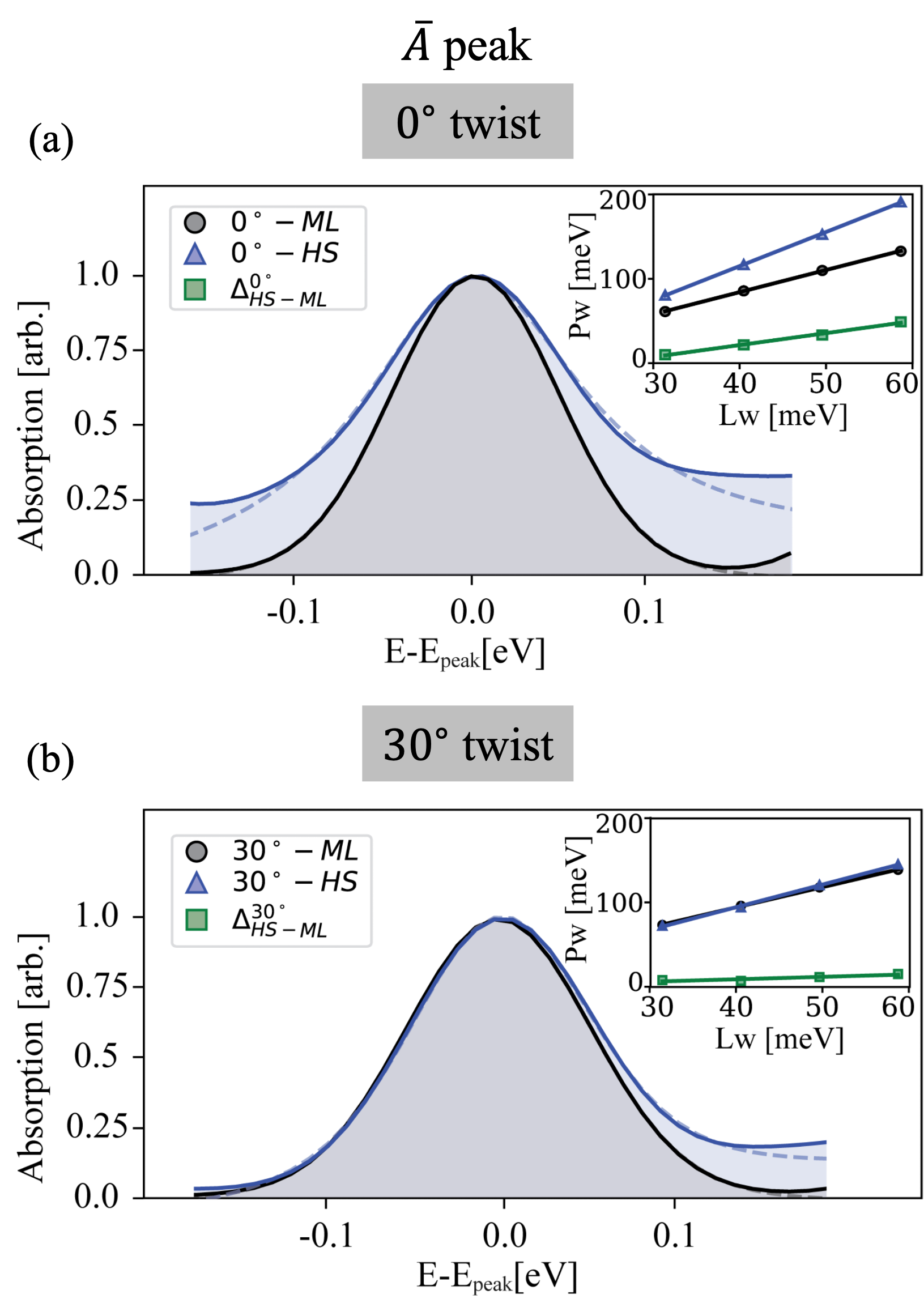}
    \caption{Spectral broadening  of the $\Tilde{\textnormal{A}}$ peak for the (a) $0^\circ$ and (b) $30^\circ$ twist HSs. Both panels compare the HS peak (blue) and the corresponding monolayer (ML) peak (black), with the corresponding fit given by dashed lines. Absorption peaks have been normalized to the peak maximum. Insets show the computed peak width (Pw) as a function of the effective Lorentzian width (Lw) of each exciton resonance for the HSs (blue triangles) and the corresponding MLs (black circles). The difference between the two, $\Delta_{HS}$ (green squares), is a manifestation of the total spectral broadening induced by structural disorder induced by graphene.}
    \label{fig:broads}
\end{figure}

A further manifestation of the exciton state hybridization and mixed nature can be observed in the absorption spectral broadening. It has been shown experimentally that the presence of the graphene layer induces an absorption line broadening compared to a separated TMD monolayer \cite{hill_exciton_2017, raja_coulomb_2017, lorchat_filtering_2020, giusca_probing_2019, tebbe_tailoring_2023}.
Our calculations allow us to computationally analyze the  broadening arising due to the above-discussed state mixing and determine their structural origins (we note that dynamical effects, expected to be dominant as well, are not accounted here). 
Our GW-BSE calculation results with a set of discrete exciton states and corresponding oscillator strengths.
In order to construct the absorption spectra shown, e.g., in Figs.~\ref{fig:optics}(a,c), we broaden each BSE exciton by a Lorentzian function centered at the exciton eigenvalue \cite{deslippe_berkeleygw_2012}, weighted by the corresponding oscillator strength, and with a scale parameter of $50$~meV to mimic the experimentally measured peak widths at room temperature
\cite{mak_atomically_2010, fang_room-temperature_2022,selig_excitonic_2016, chan_exciton_2023}.
Our above examinations reveal complex exciton compositions, with each spectral feature composed of a large number of excitations. In such a case, the actual peak width shown in the spectrum is generally larger than the effective Lorentzian broadening numerically set for each individual excitonic resonance. 
To isolate the effect of the graphene layer on the broadening, we compare the width of the $\Bar{A}$ peak in both alignments with the corresponding TMD monolayer (namely, such that includes the atomic reconstruction, but not the Gr layer), by fitting the peaks with the appropriate Voigt profiles~\cite{olivero_empirical_1977,selig_excitonic_2016} (see SI for full details).
This comparison points to strong dependence of the mixing-induced broadening on interlayer alignment. 
For the $0^\circ$-twist structure, Fig.~\ref{fig:broads}(a), the presence of graphene induces additional optically active states that significantly broaden the optical resonance. 
The overall peak width (Pw) depends linearly on the input Lorentzian width (Lw) parameter, which dictates how many transitions actually fall within the spectral peak range, as shown in Fig.~\ref{fig:broads}(c).
Importantly, the graphene-induced broadening, namely the difference between the HS and the monolayer broadening, $\Delta_{HS-ML}$, remains significant in this case and is expected to be proportional to the intrinsic TMD linewidth and at the order of 20-50 meV.
In contrast, for the $30^\circ$-twist structure, Fig.~\ref{fig:broads}(b), the presence of graphene has a negligible effect on the peak broadening (though it adds a certain degree of skewness towards higher energies). In this case, $\Delta_{HS-ML}$ is close to zero, up to $10$meV regardless of the Lorenzian scaling.
This analysis suggests that the effect of graphene on the spectral width has its origins associated with the exciton nature, and points to a direct relation between the observed broadening and the underlying structure composition and twist. More interestingly, it also suggests that the broadening difference, $\Delta_{HS-ML}$, is a parameter that captures the complexity of the many-body exciton hybridization. 

\section{Conclusion}
In conclusion, we presented a many-body first-principles study of the excitonic properties in WS\textsubscript{2}-graphene heterostructures with two different twist angles between the layers - $0^\circ$ and $30^\circ$.
We found that the optical response of these heterostructures is highly complex, with a significantly hybridized nature resulting from structure-specific mixing between interlayer and intralayer transitions. This mixing is strongly twist-angle dependent, an outcome of the relative band alignment  between the graphene Dirac cone and the TMD band edges, as well as of the momentum mismatch between the layers. 
We analyzed the nature of these transitions through the exciton charge separation and the change in absorption line broadening. 
Our results demonstrate an intriguing structural tunability of optical excitations in layered heterostructures, serving as a designable starting point for efficient relaxation in these systems. 

The data that support the findings of this study are available from the corresponding author upon reasonable request.
\newpage
\textbf{Supporting Information} \par
Supporting Information is available from the author.

\textbf{Acknowledgements} \par
Computations were carried out in the Chemfarm local cluster at the Weizmann Institute of Science and the Max Planck Computing and Data Facility cluster. A.K. and D.H.-P. acknowledge a Minerva Foundation grant 7135421. This research has received funding from the European Research Council (ERC), Grant agreement No.101041159, and from the German Research Foundation (DFG) through the Collaborative Research Center SFB 1277 (Project-ID314695032, project B10). S.R.-A.is an incumbent of the Leah Omenn Career Development Chair. 

\textbf{Conflict of interest} \par
The authors declare no conflict of interest.

\bibliography{biblio}

\begin{thebibliography}{66}%
\makeatletter
\providecommand \@ifxundefined [1]{%
 \@ifx{#1\undefined}
}%
\providecommand \@ifnum [1]{%
 \ifnum #1\expandafter \@firstoftwo
 \else \expandafter \@secondoftwo
 \fi
}%
\providecommand \@ifx [1]{%
 \ifx #1\expandafter \@firstoftwo
 \else \expandafter \@secondoftwo
 \fi
}%
\providecommand \natexlab [1]{#1}%
\providecommand \enquote  [1]{``#1''}%
\providecommand \bibnamefont  [1]{#1}%
\providecommand \bibfnamefont [1]{#1}%
\providecommand \citenamefont [1]{#1}%
\providecommand \href@noop [0]{\@secondoftwo}%
\providecommand \href [0]{\begingroup \@sanitize@url \@href}%
\providecommand \@href[1]{\@@startlink{#1}\@@href}%
\providecommand \@@href[1]{\endgroup#1\@@endlink}%
\providecommand \@sanitize@url [0]{\catcode `\\12\catcode `\$12\catcode `\&12\catcode `\#12\catcode `\^12\catcode `\_12\catcode `\%12\relax}%
\providecommand \@@startlink[1]{}%
\providecommand \@@endlink[0]{}%
\providecommand \url  [0]{\begingroup\@sanitize@url \@url }%
\providecommand \@url [1]{\endgroup\@href {#1}{\urlprefix }}%
\providecommand \urlprefix  [0]{URL }%
\providecommand \Eprint [0]{\href }%
\providecommand \doibase [0]{https://doi.org/}%
\providecommand \selectlanguage [0]{\@gobble}%
\providecommand \bibinfo  [0]{\@secondoftwo}%
\providecommand \bibfield  [0]{\@secondoftwo}%
\providecommand \translation [1]{[#1]}%
\providecommand \BibitemOpen [0]{}%
\providecommand \bibitemStop [0]{}%
\providecommand \bibitemNoStop [0]{.\EOS\space}%
\providecommand \EOS [0]{\spacefactor3000\relax}%
\providecommand \BibitemShut  [1]{\csname bibitem#1\endcsname}%
\let\auto@bib@innerbib\@empty
\bibitem [{\citenamefont {Geim}\ and\ \citenamefont {Grigorieva}(2013)}]{geim_van_2013}%
  \BibitemOpen
  \bibfield  {author} {\bibinfo {author} {\bibfnamefont {A.~K.}\ \bibnamefont {Geim}}\ and\ \bibinfo {author} {\bibfnamefont {I.~V.}\ \bibnamefont {Grigorieva}},\ }\bibfield  {title} {\bibinfo {title} {Van der {{Waals}} heterostructures},\ }\href {https://doi.org/10.1038/nature12385} {\bibfield  {journal} {\bibinfo  {journal} {Nature}\ }\textbf {\bibinfo {volume} {499}},\ \bibinfo {pages} {419} (\bibinfo {year} {2013})}\BibitemShut {NoStop}%
\bibitem [{\citenamefont {Novoselov}\ \emph {et~al.}(2016)\citenamefont {Novoselov}, \citenamefont {Mishchenko}, \citenamefont {Carvalho},\ and\ \citenamefont {Neto}}]{novoselov_2d_2016}%
  \BibitemOpen
  \bibfield  {author} {\bibinfo {author} {\bibfnamefont {K.~S.}\ \bibnamefont {Novoselov}}, \bibinfo {author} {\bibfnamefont {A.}~\bibnamefont {Mishchenko}}, \bibinfo {author} {\bibfnamefont {A.}~\bibnamefont {Carvalho}},\ and\ \bibinfo {author} {\bibfnamefont {A.~H.~C.}\ \bibnamefont {Neto}},\ }\bibfield  {title} {\bibinfo {title} {{{2D}} materials and van der {{Waals}} heterostructures},\ }\href {https://doi.org/10.1126/science.aac9439} {\bibfield  {journal} {\bibinfo  {journal} {Science}\ }\textbf {\bibinfo {volume} {353}},\ \bibinfo {pages} {aac9439} (\bibinfo {year} {2016})}\BibitemShut {NoStop}%
\bibitem [{\citenamefont {Liu}\ \emph {et~al.}(2016)\citenamefont {Liu}, \citenamefont {Weiss}, \citenamefont {Duan}, \citenamefont {Cheng}, \citenamefont {Huang},\ and\ \citenamefont {Duan}}]{liu_van_2016}%
  \BibitemOpen
  \bibfield  {author} {\bibinfo {author} {\bibfnamefont {Y.}~\bibnamefont {Liu}}, \bibinfo {author} {\bibfnamefont {N.~O.}\ \bibnamefont {Weiss}}, \bibinfo {author} {\bibfnamefont {X.}~\bibnamefont {Duan}}, \bibinfo {author} {\bibfnamefont {H.-C.}\ \bibnamefont {Cheng}}, \bibinfo {author} {\bibfnamefont {Y.}~\bibnamefont {Huang}},\ and\ \bibinfo {author} {\bibfnamefont {X.}~\bibnamefont {Duan}},\ }\bibfield  {title} {\bibinfo {title} {Van der {{Waals}} heterostructures and devices},\ }\href {https://doi.org/10.1038/natrevmats.2016.42} {\bibfield  {journal} {\bibinfo  {journal} {Nature Reviews Materials}\ }\textbf {\bibinfo {volume} {1}},\ \bibinfo {pages} {1} (\bibinfo {year} {2016})}\BibitemShut {NoStop}%
\bibitem [{\citenamefont {Massicotte}\ \emph {et~al.}(2016)\citenamefont {Massicotte}, \citenamefont {Schmidt}, \citenamefont {Vialla}, \citenamefont {Sch{\"a}dler}, \citenamefont {{Reserbat-Plantey}}, \citenamefont {Watanabe}, \citenamefont {Taniguchi}, \citenamefont {Tielrooij},\ and\ \citenamefont {Koppens}}]{massicotte_picosecond_2016}%
  \BibitemOpen
  \bibfield  {author} {\bibinfo {author} {\bibfnamefont {M.}~\bibnamefont {Massicotte}}, \bibinfo {author} {\bibfnamefont {P.}~\bibnamefont {Schmidt}}, \bibinfo {author} {\bibfnamefont {F.}~\bibnamefont {Vialla}}, \bibinfo {author} {\bibfnamefont {K.~G.}\ \bibnamefont {Sch{\"a}dler}}, \bibinfo {author} {\bibfnamefont {A.}~\bibnamefont {{Reserbat-Plantey}}}, \bibinfo {author} {\bibfnamefont {K.}~\bibnamefont {Watanabe}}, \bibinfo {author} {\bibfnamefont {T.}~\bibnamefont {Taniguchi}}, \bibinfo {author} {\bibfnamefont {K.~J.}\ \bibnamefont {Tielrooij}},\ and\ \bibinfo {author} {\bibfnamefont {F.~H.~L.}\ \bibnamefont {Koppens}},\ }\bibfield  {title} {\bibinfo {title} {Picosecond photoresponse in van der {{Waals}} heterostructures},\ }\href {https://doi.org/10.1038/nnano.2015.227} {\bibfield  {journal} {\bibinfo  {journal} {Nature Nanotechnology}\ }\textbf {\bibinfo {volume} {11}},\ \bibinfo {pages} {42} (\bibinfo {year} {2016})}\BibitemShut {NoStop}%
\bibitem [{\citenamefont {Jin}\ \emph {et~al.}(2018)\citenamefont {Jin}, \citenamefont {Ma}, \citenamefont {Karni}, \citenamefont {Regan}, \citenamefont {Wang},\ and\ \citenamefont {Heinz}}]{jin_ultrafast_2018}%
  \BibitemOpen
  \bibfield  {author} {\bibinfo {author} {\bibfnamefont {C.}~\bibnamefont {Jin}}, \bibinfo {author} {\bibfnamefont {E.~Y.}\ \bibnamefont {Ma}}, \bibinfo {author} {\bibfnamefont {O.}~\bibnamefont {Karni}}, \bibinfo {author} {\bibfnamefont {E.~C.}\ \bibnamefont {Regan}}, \bibinfo {author} {\bibfnamefont {F.}~\bibnamefont {Wang}},\ and\ \bibinfo {author} {\bibfnamefont {T.~F.}\ \bibnamefont {Heinz}},\ }\bibfield  {title} {\bibinfo {title} {Ultrafast dynamics in van der {{Waals}} heterostructures},\ }\href {https://doi.org/10.1038/s41565-018-0298-5} {\bibfield  {journal} {\bibinfo  {journal} {Nature Nanotechnology}\ }\textbf {\bibinfo {volume} {13}},\ \bibinfo {pages} {994} (\bibinfo {year} {2018})}\BibitemShut {NoStop}%
\bibitem [{\citenamefont {Kennes}\ \emph {et~al.}(2021)\citenamefont {Kennes}, \citenamefont {Claassen}, \citenamefont {Xian}, \citenamefont {Georges}, \citenamefont {Millis}, \citenamefont {Hone}, \citenamefont {Dean}, \citenamefont {Basov}, \citenamefont {Pasupathy},\ and\ \citenamefont {Rubio}}]{kennes_moire_2021}%
  \BibitemOpen
  \bibfield  {author} {\bibinfo {author} {\bibfnamefont {D.~M.}\ \bibnamefont {Kennes}}, \bibinfo {author} {\bibfnamefont {M.}~\bibnamefont {Claassen}}, \bibinfo {author} {\bibfnamefont {L.}~\bibnamefont {Xian}}, \bibinfo {author} {\bibfnamefont {A.}~\bibnamefont {Georges}}, \bibinfo {author} {\bibfnamefont {A.~J.}\ \bibnamefont {Millis}}, \bibinfo {author} {\bibfnamefont {J.}~\bibnamefont {Hone}}, \bibinfo {author} {\bibfnamefont {C.~R.}\ \bibnamefont {Dean}}, \bibinfo {author} {\bibfnamefont {D.~N.}\ \bibnamefont {Basov}}, \bibinfo {author} {\bibfnamefont {A.~N.}\ \bibnamefont {Pasupathy}},\ and\ \bibinfo {author} {\bibfnamefont {A.}~\bibnamefont {Rubio}},\ }\bibfield  {title} {\bibinfo {title} {Moir{\'e} heterostructures as a condensed-matter quantum simulator},\ }\href {https://doi.org/10.1038/s41567-020-01154-3} {\bibfield  {journal} {\bibinfo  {journal} {Nature Physics}\ }\textbf {\bibinfo {volume} {17}},\ \bibinfo {pages} {155} (\bibinfo {year} {2021})}\BibitemShut {NoStop}%
\bibitem [{\citenamefont {Lemme}\ \emph {et~al.}(2022)\citenamefont {Lemme}, \citenamefont {Akinwande}, \citenamefont {Huyghebaert},\ and\ \citenamefont {Stampfer}}]{lemme_2d_2022}%
  \BibitemOpen
  \bibfield  {author} {\bibinfo {author} {\bibfnamefont {M.~C.}\ \bibnamefont {Lemme}}, \bibinfo {author} {\bibfnamefont {D.}~\bibnamefont {Akinwande}}, \bibinfo {author} {\bibfnamefont {C.}~\bibnamefont {Huyghebaert}},\ and\ \bibinfo {author} {\bibfnamefont {C.}~\bibnamefont {Stampfer}},\ }\bibfield  {title} {\bibinfo {title} {{{2D}} materials for future heterogeneous electronics},\ }\href {https://doi.org/10.1038/s41467-022-29001-4} {\bibfield  {journal} {\bibinfo  {journal} {Nature Communications}\ }\textbf {\bibinfo {volume} {13}},\ \bibinfo {pages} {1392} (\bibinfo {year} {2022})}\BibitemShut {NoStop}%
\bibitem [{\citenamefont {Jin}\ \emph {et~al.}(2015)\citenamefont {Jin}, \citenamefont {Rasmussen},\ and\ \citenamefont {Thygesen}}]{jin_tuning_2015}%
  \BibitemOpen
  \bibfield  {author} {\bibinfo {author} {\bibfnamefont {C.}~\bibnamefont {Jin}}, \bibinfo {author} {\bibfnamefont {F.~A.}\ \bibnamefont {Rasmussen}},\ and\ \bibinfo {author} {\bibfnamefont {K.~S.}\ \bibnamefont {Thygesen}},\ }\bibfield  {title} {\bibinfo {title} {Tuning the {{Schottky Barrier}} at the {{Graphene}}/{{MoS2 Interface}} by {{Electron Doping}}: {{Density Functional Theory}} and {{Many-Body Calculations}}},\ }\href {https://doi.org/10.1021/acs.jpcc.5b05580} {\bibfield  {journal} {\bibinfo  {journal} {The Journal of Physical Chemistry C}\ }\textbf {\bibinfo {volume} {119}},\ \bibinfo {pages} {19928} (\bibinfo {year} {2015})}\BibitemShut {NoStop}%
\bibitem [{\citenamefont {Hill}\ \emph {et~al.}(2017)\citenamefont {Hill}, \citenamefont {Rigosi}, \citenamefont {Raja}, \citenamefont {Chernikov}, \citenamefont {Roquelet},\ and\ \citenamefont {Heinz}}]{hill_exciton_2017}%
  \BibitemOpen
  \bibfield  {author} {\bibinfo {author} {\bibfnamefont {H.~M.}\ \bibnamefont {Hill}}, \bibinfo {author} {\bibfnamefont {A.~F.}\ \bibnamefont {Rigosi}}, \bibinfo {author} {\bibfnamefont {A.}~\bibnamefont {Raja}}, \bibinfo {author} {\bibfnamefont {A.}~\bibnamefont {Chernikov}}, \bibinfo {author} {\bibfnamefont {C.}~\bibnamefont {Roquelet}},\ and\ \bibinfo {author} {\bibfnamefont {T.~F.}\ \bibnamefont {Heinz}},\ }\bibfield  {title} {\bibinfo {title} {Exciton broadening in {WS2}/ graphene heterostructures},\ }\href {https://doi.org/10.1103/PhysRevB.96.205401} {\bibfield  {journal} {\bibinfo  {journal} {Physical Review B}\ }\textbf {\bibinfo {volume} {96}},\ \bibinfo {pages} {205401} (\bibinfo {year} {2017})}\BibitemShut {NoStop}%
\bibitem [{\citenamefont {Aeschlimann}\ \emph {et~al.}(2020)\citenamefont {Aeschlimann}, \citenamefont {Rossi}, \citenamefont {{Ch{\'a}vez-Cervantes}}, \citenamefont {Krause}, \citenamefont {Arnoldi}, \citenamefont {Stadtm{\"u}ller}, \citenamefont {Aeschlimann}, \citenamefont {Forti}, \citenamefont {Fabbri}, \citenamefont {Coletti},\ and\ \citenamefont {Gierz}}]{aeschlimann_direct_2020}%
  \BibitemOpen
  \bibfield  {author} {\bibinfo {author} {\bibfnamefont {S.}~\bibnamefont {Aeschlimann}}, \bibinfo {author} {\bibfnamefont {A.}~\bibnamefont {Rossi}}, \bibinfo {author} {\bibfnamefont {M.}~\bibnamefont {{Ch{\'a}vez-Cervantes}}}, \bibinfo {author} {\bibfnamefont {R.}~\bibnamefont {Krause}}, \bibinfo {author} {\bibfnamefont {B.}~\bibnamefont {Arnoldi}}, \bibinfo {author} {\bibfnamefont {B.}~\bibnamefont {Stadtm{\"u}ller}}, \bibinfo {author} {\bibfnamefont {M.}~\bibnamefont {Aeschlimann}}, \bibinfo {author} {\bibfnamefont {S.}~\bibnamefont {Forti}}, \bibinfo {author} {\bibfnamefont {F.}~\bibnamefont {Fabbri}}, \bibinfo {author} {\bibfnamefont {C.}~\bibnamefont {Coletti}},\ and\ \bibinfo {author} {\bibfnamefont {I.}~\bibnamefont {Gierz}},\ }\bibfield  {title} {\bibinfo {title} {Direct evidence for efficient ultrafast charge separation in epitaxial {{WS2}}/graphene heterostructures},\ }\href {https://doi.org/10.1126/sciadv.aay0761} {\bibfield  {journal} {\bibinfo  {journal} {Science Advances}\ }\textbf {\bibinfo
  {volume} {6}},\ \bibinfo {pages} {eaay0761} (\bibinfo {year} {2020})}\BibitemShut {NoStop}%
\bibitem [{\citenamefont {Krause}\ \emph {et~al.}(2021{\natexlab{a}})\citenamefont {Krause}, \citenamefont {Aeschlimann}, \citenamefont {{Ch{\'a}vez-Cervantes}}, \citenamefont {{Perea-Causin}}, \citenamefont {Brem}, \citenamefont {Malic}, \citenamefont {Forti}, \citenamefont {Fabbri}, \citenamefont {Coletti},\ and\ \citenamefont {Gierz}}]{krause_microscopic_2021}%
  \BibitemOpen
  \bibfield  {author} {\bibinfo {author} {\bibfnamefont {R.}~\bibnamefont {Krause}}, \bibinfo {author} {\bibfnamefont {S.}~\bibnamefont {Aeschlimann}}, \bibinfo {author} {\bibfnamefont {M.}~\bibnamefont {{Ch{\'a}vez-Cervantes}}}, \bibinfo {author} {\bibfnamefont {R.}~\bibnamefont {{Perea-Causin}}}, \bibinfo {author} {\bibfnamefont {S.}~\bibnamefont {Brem}}, \bibinfo {author} {\bibfnamefont {E.}~\bibnamefont {Malic}}, \bibinfo {author} {\bibfnamefont {S.}~\bibnamefont {Forti}}, \bibinfo {author} {\bibfnamefont {F.}~\bibnamefont {Fabbri}}, \bibinfo {author} {\bibfnamefont {C.}~\bibnamefont {Coletti}},\ and\ \bibinfo {author} {\bibfnamefont {I.}~\bibnamefont {Gierz}},\ }\bibfield  {title} {\bibinfo {title} {Microscopic {{Understanding}} of {{Ultrafast Charge Transfer}} in van der {{Waals Heterostructures}}},\ }\href {https://doi.org/10.1103/PhysRevLett.127.276401} {\bibfield  {journal} {\bibinfo  {journal} {Physical Review Letters}\ }\textbf {\bibinfo {volume} {127}},\ \bibinfo {pages} {276401} (\bibinfo {year}
  {2021}{\natexlab{a}})}\BibitemShut {NoStop}%
\bibitem [{\citenamefont {Novoselov}\ \emph {et~al.}(2004)\citenamefont {Novoselov}, \citenamefont {Geim}, \citenamefont {Morozov}, \citenamefont {Jiang}, \citenamefont {Zhang}, \citenamefont {Dubonos}, \citenamefont {Grigorieva},\ and\ \citenamefont {Firsov}}]{novoselov_electric_2004}%
  \BibitemOpen
  \bibfield  {author} {\bibinfo {author} {\bibfnamefont {K.~S.}\ \bibnamefont {Novoselov}}, \bibinfo {author} {\bibfnamefont {A.~K.}\ \bibnamefont {Geim}}, \bibinfo {author} {\bibfnamefont {S.~V.}\ \bibnamefont {Morozov}}, \bibinfo {author} {\bibfnamefont {D.}~\bibnamefont {Jiang}}, \bibinfo {author} {\bibfnamefont {Y.}~\bibnamefont {Zhang}}, \bibinfo {author} {\bibfnamefont {S.~V.}\ \bibnamefont {Dubonos}}, \bibinfo {author} {\bibfnamefont {I.~V.}\ \bibnamefont {Grigorieva}},\ and\ \bibinfo {author} {\bibfnamefont {A.~A.}\ \bibnamefont {Firsov}},\ }\bibfield  {title} {\bibinfo {title} {Electric {{Field Effect}} in {{Atomically Thin Carbon Films}}},\ }\href {https://doi.org/10.1126/science.1102896} {\bibfield  {journal} {\bibinfo  {journal} {Science}\ }\textbf {\bibinfo {volume} {306}},\ \bibinfo {pages} {666} (\bibinfo {year} {2004})}\BibitemShut {NoStop}%
\bibitem [{\citenamefont {Castro~Neto}\ \emph {et~al.}(2009)\citenamefont {Castro~Neto}, \citenamefont {Guinea}, \citenamefont {Peres}, \citenamefont {Novoselov},\ and\ \citenamefont {Geim}}]{castro_neto_electronic_2009}%
  \BibitemOpen
  \bibfield  {author} {\bibinfo {author} {\bibfnamefont {A.~H.}\ \bibnamefont {Castro~Neto}}, \bibinfo {author} {\bibfnamefont {F.}~\bibnamefont {Guinea}}, \bibinfo {author} {\bibfnamefont {N.~M.~R.}\ \bibnamefont {Peres}}, \bibinfo {author} {\bibfnamefont {K.~S.}\ \bibnamefont {Novoselov}},\ and\ \bibinfo {author} {\bibfnamefont {A.~K.}\ \bibnamefont {Geim}},\ }\bibfield  {title} {\bibinfo {title} {The electronic properties of graphene},\ }\href {https://doi.org/10.1103/RevModPhys.81.109} {\bibfield  {journal} {\bibinfo  {journal} {Reviews of Modern Physics}\ }\textbf {\bibinfo {volume} {81}},\ \bibinfo {pages} {109} (\bibinfo {year} {2009})}\BibitemShut {NoStop}%
\bibitem [{\citenamefont {Yang}\ \emph {et~al.}(2009)\citenamefont {Yang}, \citenamefont {Deslippe}, \citenamefont {Park}, \citenamefont {Cohen},\ and\ \citenamefont {Louie}}]{yang_excitonic_2009}%
  \BibitemOpen
  \bibfield  {author} {\bibinfo {author} {\bibfnamefont {L.}~\bibnamefont {Yang}}, \bibinfo {author} {\bibfnamefont {J.}~\bibnamefont {Deslippe}}, \bibinfo {author} {\bibfnamefont {C.-H.}\ \bibnamefont {Park}}, \bibinfo {author} {\bibfnamefont {M.~L.}\ \bibnamefont {Cohen}},\ and\ \bibinfo {author} {\bibfnamefont {S.~G.}\ \bibnamefont {Louie}},\ }\bibfield  {title} {\bibinfo {title} {Excitonic {{Effects}} on the {{Optical Response}} of {{Graphene}} and {{Bilayer Graphene}}},\ }\href {https://doi.org/10.1103/PhysRevLett.103.186802} {\bibfield  {journal} {\bibinfo  {journal} {Physical Review Letters}\ }\textbf {\bibinfo {volume} {103}},\ \bibinfo {pages} {186802} (\bibinfo {year} {2009})}\BibitemShut {NoStop}%
\bibitem [{\citenamefont {Mak}\ \emph {et~al.}(2008)\citenamefont {Mak}, \citenamefont {Sfeir}, \citenamefont {Wu}, \citenamefont {Lui}, \citenamefont {Misewich},\ and\ \citenamefont {Heinz}}]{mak_measurement_2008}%
  \BibitemOpen
  \bibfield  {author} {\bibinfo {author} {\bibfnamefont {K.~F.}\ \bibnamefont {Mak}}, \bibinfo {author} {\bibfnamefont {M.~Y.}\ \bibnamefont {Sfeir}}, \bibinfo {author} {\bibfnamefont {Y.}~\bibnamefont {Wu}}, \bibinfo {author} {\bibfnamefont {C.~H.}\ \bibnamefont {Lui}}, \bibinfo {author} {\bibfnamefont {J.~A.}\ \bibnamefont {Misewich}},\ and\ \bibinfo {author} {\bibfnamefont {T.~F.}\ \bibnamefont {Heinz}},\ }\bibfield  {title} {\bibinfo {title} {Measurement of the {{Optical Conductivity}} of {{Graphene}}},\ }\href {https://doi.org/10.1103/PhysRevLett.101.196405} {\bibfield  {journal} {\bibinfo  {journal} {Physical Review Letters}\ }\textbf {\bibinfo {volume} {101}},\ \bibinfo {pages} {196405} (\bibinfo {year} {2008})}\BibitemShut {NoStop}%
\bibitem [{\citenamefont {Das~Sarma}\ \emph {et~al.}(2011)\citenamefont {Das~Sarma}, \citenamefont {Adam}, \citenamefont {Hwang},\ and\ \citenamefont {Rossi}}]{das_sarma_electronic_2011}%
  \BibitemOpen
  \bibfield  {author} {\bibinfo {author} {\bibfnamefont {S.}~\bibnamefont {Das~Sarma}}, \bibinfo {author} {\bibfnamefont {S.}~\bibnamefont {Adam}}, \bibinfo {author} {\bibfnamefont {E.~H.}\ \bibnamefont {Hwang}},\ and\ \bibinfo {author} {\bibfnamefont {E.}~\bibnamefont {Rossi}},\ }\bibfield  {title} {\bibinfo {title} {Electronic transport in two-dimensional graphene},\ }\href {https://doi.org/10.1103/RevModPhys.83.407} {\bibfield  {journal} {\bibinfo  {journal} {Reviews of Modern Physics}\ }\textbf {\bibinfo {volume} {83}},\ \bibinfo {pages} {407} (\bibinfo {year} {2011})}\BibitemShut {NoStop}%
\bibitem [{\citenamefont {Mak}\ \emph {et~al.}(2014)\citenamefont {Mak}, \citenamefont {Da~Jornada}, \citenamefont {He}, \citenamefont {Deslippe}, \citenamefont {Petrone}, \citenamefont {Hone}, \citenamefont {Shan}, \citenamefont {Louie},\ and\ \citenamefont {Heinz}}]{mak2014tuning}%
  \BibitemOpen
  \bibfield  {author} {\bibinfo {author} {\bibfnamefont {K.~F.}\ \bibnamefont {Mak}}, \bibinfo {author} {\bibfnamefont {F.~H.}\ \bibnamefont {Da~Jornada}}, \bibinfo {author} {\bibfnamefont {K.}~\bibnamefont {He}}, \bibinfo {author} {\bibfnamefont {J.}~\bibnamefont {Deslippe}}, \bibinfo {author} {\bibfnamefont {N.}~\bibnamefont {Petrone}}, \bibinfo {author} {\bibfnamefont {J.}~\bibnamefont {Hone}}, \bibinfo {author} {\bibfnamefont {J.}~\bibnamefont {Shan}}, \bibinfo {author} {\bibfnamefont {S.~G.}\ \bibnamefont {Louie}},\ and\ \bibinfo {author} {\bibfnamefont {T.~F.}\ \bibnamefont {Heinz}},\ }\bibfield  {title} {\bibinfo {title} {Tuning many-body interactions in graphene: The effects of doping on excitons and carrier lifetimes},\ }\href@noop {} {\bibfield  {journal} {\bibinfo  {journal} {Physical review letters}\ }\textbf {\bibinfo {volume} {112}},\ \bibinfo {pages} {207401} (\bibinfo {year} {2014})}\BibitemShut {NoStop}%
\bibitem [{\citenamefont {Mak}\ \emph {et~al.}(2010)\citenamefont {Mak}, \citenamefont {Lee}, \citenamefont {Hone}, \citenamefont {Shan},\ and\ \citenamefont {Heinz}}]{mak_atomically_2010}%
  \BibitemOpen
  \bibfield  {author} {\bibinfo {author} {\bibfnamefont {K.~F.}\ \bibnamefont {Mak}}, \bibinfo {author} {\bibfnamefont {C.}~\bibnamefont {Lee}}, \bibinfo {author} {\bibfnamefont {J.}~\bibnamefont {Hone}}, \bibinfo {author} {\bibfnamefont {J.}~\bibnamefont {Shan}},\ and\ \bibinfo {author} {\bibfnamefont {T.~F.}\ \bibnamefont {Heinz}},\ }\bibfield  {title} {\bibinfo {title} {Atomically {{Thin MoS2}}: {{A New Direct-Gap Semiconductor}}},\ }\href {https://doi.org/10.1103/PhysRevLett.105.136805} {\bibfield  {journal} {\bibinfo  {journal} {Physical Review Letters}\ }\textbf {\bibinfo {volume} {105}},\ \bibinfo {pages} {136805} (\bibinfo {year} {2010})}\BibitemShut {NoStop}%
\bibitem [{\citenamefont {Splendiani}\ \emph {et~al.}(2010)\citenamefont {Splendiani}, \citenamefont {Sun}, \citenamefont {Zhang}, \citenamefont {Li}, \citenamefont {Kim}, \citenamefont {Chim}, \citenamefont {Galli},\ and\ \citenamefont {Wang}}]{splendiani_emerging_2010}%
  \BibitemOpen
  \bibfield  {author} {\bibinfo {author} {\bibfnamefont {A.}~\bibnamefont {Splendiani}}, \bibinfo {author} {\bibfnamefont {L.}~\bibnamefont {Sun}}, \bibinfo {author} {\bibfnamefont {Y.}~\bibnamefont {Zhang}}, \bibinfo {author} {\bibfnamefont {T.}~\bibnamefont {Li}}, \bibinfo {author} {\bibfnamefont {J.}~\bibnamefont {Kim}}, \bibinfo {author} {\bibfnamefont {C.-Y.}\ \bibnamefont {Chim}}, \bibinfo {author} {\bibfnamefont {G.}~\bibnamefont {Galli}},\ and\ \bibinfo {author} {\bibfnamefont {F.}~\bibnamefont {Wang}},\ }\bibfield  {title} {\bibinfo {title} {Emerging {{Photoluminescence}} in {{Monolayer MoS2}}},\ }\href {https://doi.org/10.1021/nl903868w} {\bibfield  {journal} {\bibinfo  {journal} {Nano Letters}\ }\textbf {\bibinfo {volume} {10}},\ \bibinfo {pages} {1271} (\bibinfo {year} {2010})}\BibitemShut {NoStop}%
\bibitem [{\citenamefont {Ramasubramaniam}(2012)}]{ramasubramaniam_large_2012}%
  \BibitemOpen
  \bibfield  {author} {\bibinfo {author} {\bibfnamefont {A.}~\bibnamefont {Ramasubramaniam}},\ }\bibfield  {title} {\bibinfo {title} {Large excitonic effects in monolayers of molybdenum and tungsten dichalcogenides},\ }\href {https://doi.org/10.1103/PhysRevB.86.115409} {\bibfield  {journal} {\bibinfo  {journal} {Physical Review B}\ }\textbf {\bibinfo {volume} {86}},\ \bibinfo {pages} {115409} (\bibinfo {year} {2012})}\BibitemShut {NoStop}%
\bibitem [{\citenamefont {Ugeda}\ \emph {et~al.}(2014)\citenamefont {Ugeda}, \citenamefont {Bradley}, \citenamefont {Shi}, \citenamefont {{da Jornada}}, \citenamefont {Zhang}, \citenamefont {Qiu}, \citenamefont {Ruan}, \citenamefont {Mo}, \citenamefont {Hussain}, \citenamefont {Shen}, \citenamefont {Wang}, \citenamefont {Louie},\ and\ \citenamefont {Crommie}}]{ugeda_giant_2014}%
  \BibitemOpen
  \bibfield  {author} {\bibinfo {author} {\bibfnamefont {M.~M.}\ \bibnamefont {Ugeda}}, \bibinfo {author} {\bibfnamefont {A.~J.}\ \bibnamefont {Bradley}}, \bibinfo {author} {\bibfnamefont {S.-F.}\ \bibnamefont {Shi}}, \bibinfo {author} {\bibfnamefont {F.~H.}\ \bibnamefont {{da Jornada}}}, \bibinfo {author} {\bibfnamefont {Y.}~\bibnamefont {Zhang}}, \bibinfo {author} {\bibfnamefont {D.~Y.}\ \bibnamefont {Qiu}}, \bibinfo {author} {\bibfnamefont {W.}~\bibnamefont {Ruan}}, \bibinfo {author} {\bibfnamefont {S.-K.}\ \bibnamefont {Mo}}, \bibinfo {author} {\bibfnamefont {Z.}~\bibnamefont {Hussain}}, \bibinfo {author} {\bibfnamefont {Z.-X.}\ \bibnamefont {Shen}}, \bibinfo {author} {\bibfnamefont {F.}~\bibnamefont {Wang}}, \bibinfo {author} {\bibfnamefont {S.~G.}\ \bibnamefont {Louie}},\ and\ \bibinfo {author} {\bibfnamefont {M.~F.}\ \bibnamefont {Crommie}},\ }\bibfield  {title} {\bibinfo {title} {Giant bandgap renormalization and excitonic effects in a monolayer transition metal dichalcogenide semiconductor},\ }\href
  {https://doi.org/10.1038/nmat4061} {\bibfield  {journal} {\bibinfo  {journal} {Nature Materials}\ }\textbf {\bibinfo {volume} {13}},\ \bibinfo {pages} {1091} (\bibinfo {year} {2014})}\BibitemShut {NoStop}%
\bibitem [{\citenamefont {Chernikov}\ \emph {et~al.}(2014)\citenamefont {Chernikov}, \citenamefont {Berkelbach}, \citenamefont {Hill}, \citenamefont {Rigosi}, \citenamefont {Li}, \citenamefont {Aslan}, \citenamefont {Reichman}, \citenamefont {Hybertsen},\ and\ \citenamefont {Heinz}}]{chernikov_exciton_2014}%
  \BibitemOpen
  \bibfield  {author} {\bibinfo {author} {\bibfnamefont {A.}~\bibnamefont {Chernikov}}, \bibinfo {author} {\bibfnamefont {T.~C.}\ \bibnamefont {Berkelbach}}, \bibinfo {author} {\bibfnamefont {H.~M.}\ \bibnamefont {Hill}}, \bibinfo {author} {\bibfnamefont {A.}~\bibnamefont {Rigosi}}, \bibinfo {author} {\bibfnamefont {Y.}~\bibnamefont {Li}}, \bibinfo {author} {\bibfnamefont {O.~B.}\ \bibnamefont {Aslan}}, \bibinfo {author} {\bibfnamefont {D.~R.}\ \bibnamefont {Reichman}}, \bibinfo {author} {\bibfnamefont {M.~S.}\ \bibnamefont {Hybertsen}},\ and\ \bibinfo {author} {\bibfnamefont {T.~F.}\ \bibnamefont {Heinz}},\ }\bibfield  {title} {\bibinfo {title} {Exciton {{Binding Energy}} and {{Nonhydrogenic Rydberg Series}} in {{Monolayer WS2}}},\ }\href {https://doi.org/10.1103/PhysRevLett.113.076802} {\bibfield  {journal} {\bibinfo  {journal} {Physical Review Letters}\ }\textbf {\bibinfo {volume} {113}},\ \bibinfo {pages} {076802} (\bibinfo {year} {2014})}\BibitemShut {NoStop}%
\bibitem [{\citenamefont {Hanbicki}\ \emph {et~al.}(2015)\citenamefont {Hanbicki}, \citenamefont {Currie}, \citenamefont {Kioseoglou}, \citenamefont {Friedman},\ and\ \citenamefont {Jonker}}]{hanbicki_measurement_2015}%
  \BibitemOpen
  \bibfield  {author} {\bibinfo {author} {\bibfnamefont {A.~T.}\ \bibnamefont {Hanbicki}}, \bibinfo {author} {\bibfnamefont {M.}~\bibnamefont {Currie}}, \bibinfo {author} {\bibfnamefont {G.}~\bibnamefont {Kioseoglou}}, \bibinfo {author} {\bibfnamefont {A.~L.}\ \bibnamefont {Friedman}},\ and\ \bibinfo {author} {\bibfnamefont {B.~T.}\ \bibnamefont {Jonker}},\ }\bibfield  {title} {\bibinfo {title} {Measurement of high exciton binding energy in the monolayer transition-metal dichalcogenides {{WS2}} and {{WSe2}}},\ }\href {https://doi.org/10.1016/j.ssc.2014.11.005} {\bibfield  {journal} {\bibinfo  {journal} {Solid State Communications}\ }\textbf {\bibinfo {volume} {203}},\ \bibinfo {pages} {16} (\bibinfo {year} {2015})}\BibitemShut {NoStop}%
\bibitem [{\citenamefont {Song}\ \emph {et~al.}(2018)\citenamefont {Song}, \citenamefont {Zhu}, \citenamefont {Shi}, \citenamefont {Sun},\ and\ \citenamefont {Ruan}}]{song_ultrafast_2018}%
  \BibitemOpen
  \bibfield  {author} {\bibinfo {author} {\bibfnamefont {Z.}~\bibnamefont {Song}}, \bibinfo {author} {\bibfnamefont {H.}~\bibnamefont {Zhu}}, \bibinfo {author} {\bibfnamefont {W.}~\bibnamefont {Shi}}, \bibinfo {author} {\bibfnamefont {D.}~\bibnamefont {Sun}},\ and\ \bibinfo {author} {\bibfnamefont {S.}~\bibnamefont {Ruan}},\ }\bibfield  {title} {\bibinfo {title} {Ultrafast charge transfer in graphene-{{WS2 Van}} der {{Waals}} heterostructures},\ }\href {https://doi.org/10.1016/j.ijleo.2018.08.059} {\bibfield  {journal} {\bibinfo  {journal} {Optik}\ }\textbf {\bibinfo {volume} {174}},\ \bibinfo {pages} {62} (\bibinfo {year} {2018})}\BibitemShut {NoStop}%
\bibitem [{\citenamefont {Krause}\ \emph {et~al.}(2021{\natexlab{b}})\citenamefont {Krause}, \citenamefont {{Ch{\'a}vez-Cervantes}}, \citenamefont {Aeschlimann}, \citenamefont {Forti}, \citenamefont {Fabbri}, \citenamefont {Rossi}, \citenamefont {Coletti}, \citenamefont {Cacho}, \citenamefont {Zhang}, \citenamefont {Majchrzak}, \citenamefont {Chapman}, \citenamefont {Springate},\ and\ \citenamefont {Gierz}}]{krause_ultrafast_2021}%
  \BibitemOpen
  \bibfield  {author} {\bibinfo {author} {\bibfnamefont {R.}~\bibnamefont {Krause}}, \bibinfo {author} {\bibfnamefont {M.}~\bibnamefont {{Ch{\'a}vez-Cervantes}}}, \bibinfo {author} {\bibfnamefont {S.}~\bibnamefont {Aeschlimann}}, \bibinfo {author} {\bibfnamefont {S.}~\bibnamefont {Forti}}, \bibinfo {author} {\bibfnamefont {F.}~\bibnamefont {Fabbri}}, \bibinfo {author} {\bibfnamefont {A.}~\bibnamefont {Rossi}}, \bibinfo {author} {\bibfnamefont {C.}~\bibnamefont {Coletti}}, \bibinfo {author} {\bibfnamefont {C.}~\bibnamefont {Cacho}}, \bibinfo {author} {\bibfnamefont {Y.}~\bibnamefont {Zhang}}, \bibinfo {author} {\bibfnamefont {P.~E.}\ \bibnamefont {Majchrzak}}, \bibinfo {author} {\bibfnamefont {R.~T.}\ \bibnamefont {Chapman}}, \bibinfo {author} {\bibfnamefont {E.}~\bibnamefont {Springate}},\ and\ \bibinfo {author} {\bibfnamefont {I.}~\bibnamefont {Gierz}},\ }\bibfield  {title} {\bibinfo {title} {Ultrafast {{Charge Separation}} in {{Bilayer WS2}}/{{Graphene Heterostructure Revealed}} by {{Time-}} and
  {{Angle-Resolved Photoemission Spectroscopy}}},\ }\href@noop {} {\bibfield  {journal} {\bibinfo  {journal} {Frontiers in Physics}\ }\textbf {\bibinfo {volume} {9}} (\bibinfo {year} {2021}{\natexlab{b}})}\BibitemShut {NoStop}%
\bibitem [{\citenamefont {Trovatello}\ \emph {et~al.}(2022)\citenamefont {Trovatello}, \citenamefont {Piccinini}, \citenamefont {Forti}, \citenamefont {Fabbri}, \citenamefont {Rossi}, \citenamefont {De~Silvestri}, \citenamefont {Coletti}, \citenamefont {Cerullo},\ and\ \citenamefont {Dal~Conte}}]{trovatello_ultrafast_2022}%
  \BibitemOpen
  \bibfield  {author} {\bibinfo {author} {\bibfnamefont {C.}~\bibnamefont {Trovatello}}, \bibinfo {author} {\bibfnamefont {G.}~\bibnamefont {Piccinini}}, \bibinfo {author} {\bibfnamefont {S.}~\bibnamefont {Forti}}, \bibinfo {author} {\bibfnamefont {F.}~\bibnamefont {Fabbri}}, \bibinfo {author} {\bibfnamefont {A.}~\bibnamefont {Rossi}}, \bibinfo {author} {\bibfnamefont {S.}~\bibnamefont {De~Silvestri}}, \bibinfo {author} {\bibfnamefont {C.}~\bibnamefont {Coletti}}, \bibinfo {author} {\bibfnamefont {G.}~\bibnamefont {Cerullo}},\ and\ \bibinfo {author} {\bibfnamefont {S.}~\bibnamefont {Dal~Conte}},\ }\bibfield  {title} {\bibinfo {title} {Ultrafast hot carrier transfer in {{WS2}}/graphene large area heterostructures},\ }\href {https://doi.org/10.1038/s41699-022-00299-4} {\bibfield  {journal} {\bibinfo  {journal} {npj 2D Materials and Applications}\ }\textbf {\bibinfo {volume} {6}},\ \bibinfo {pages} {1} (\bibinfo {year} {2022})}\BibitemShut {NoStop}%
\bibitem [{\citenamefont {Fu}\ \emph {et~al.}(2021)\citenamefont {Fu}, \citenamefont {{du Foss{\'e}}}, \citenamefont {Jia}, \citenamefont {Xu}, \citenamefont {Yu}, \citenamefont {Zhang}, \citenamefont {Zheng}, \citenamefont {Krasel}, \citenamefont {Chen}, \citenamefont {Wang}, \citenamefont {Tielrooij}, \citenamefont {Bonn}, \citenamefont {Houtepen},\ and\ \citenamefont {Wang}}]{fu_long-lived_2021}%
  \BibitemOpen
  \bibfield  {author} {\bibinfo {author} {\bibfnamefont {S.}~\bibnamefont {Fu}}, \bibinfo {author} {\bibfnamefont {I.}~\bibnamefont {{du Foss{\'e}}}}, \bibinfo {author} {\bibfnamefont {X.}~\bibnamefont {Jia}}, \bibinfo {author} {\bibfnamefont {J.}~\bibnamefont {Xu}}, \bibinfo {author} {\bibfnamefont {X.}~\bibnamefont {Yu}}, \bibinfo {author} {\bibfnamefont {H.}~\bibnamefont {Zhang}}, \bibinfo {author} {\bibfnamefont {W.}~\bibnamefont {Zheng}}, \bibinfo {author} {\bibfnamefont {S.}~\bibnamefont {Krasel}}, \bibinfo {author} {\bibfnamefont {Z.}~\bibnamefont {Chen}}, \bibinfo {author} {\bibfnamefont {Z.~M.}\ \bibnamefont {Wang}}, \bibinfo {author} {\bibfnamefont {K.-J.}\ \bibnamefont {Tielrooij}}, \bibinfo {author} {\bibfnamefont {M.}~\bibnamefont {Bonn}}, \bibinfo {author} {\bibfnamefont {A.~J.}\ \bibnamefont {Houtepen}},\ and\ \bibinfo {author} {\bibfnamefont {H.~I.}\ \bibnamefont {Wang}},\ }\bibfield  {title} {\bibinfo {title} {Long-lived charge separation following pump-wavelength{\textendash}dependent
  ultrafast charge transfer in graphene/{{WS2}} heterostructures},\ }\href {https://doi.org/10.1126/sciadv.abd9061} {\bibfield  {journal} {\bibinfo  {journal} {Science Advances}\ }\textbf {\bibinfo {volume} {7}},\ \bibinfo {pages} {eabd9061} (\bibinfo {year} {2021})}\BibitemShut {NoStop}%
\bibitem [{\citenamefont {Dong}\ \emph {et~al.}(2023)\citenamefont {Dong}, \citenamefont {Beaulieu}, \citenamefont {Selig}, \citenamefont {Rosenzweig}, \citenamefont {Christiansen}, \citenamefont {Pincelli}, \citenamefont {Dendzik}, \citenamefont {Ziegler}, \citenamefont {Maklar}, \citenamefont {Xian}, \citenamefont {Neef}, \citenamefont {Mohammed}, \citenamefont {Schulz}, \citenamefont {Stadler}, \citenamefont {Jetter}, \citenamefont {Michler}, \citenamefont {Taniguchi}, \citenamefont {Watanabe}, \citenamefont {Takagi}, \citenamefont {Starke}, \citenamefont {Chernikov}, \citenamefont {Wolf}, \citenamefont {Nakamura}, \citenamefont {Knorr}, \citenamefont {Rettig},\ and\ \citenamefont {Ernstorfer}}]{dong_observation_2023}%
  \BibitemOpen
  \bibfield  {author} {\bibinfo {author} {\bibfnamefont {S.}~\bibnamefont {Dong}}, \bibinfo {author} {\bibfnamefont {S.}~\bibnamefont {Beaulieu}}, \bibinfo {author} {\bibfnamefont {M.}~\bibnamefont {Selig}}, \bibinfo {author} {\bibfnamefont {P.}~\bibnamefont {Rosenzweig}}, \bibinfo {author} {\bibfnamefont {D.}~\bibnamefont {Christiansen}}, \bibinfo {author} {\bibfnamefont {T.}~\bibnamefont {Pincelli}}, \bibinfo {author} {\bibfnamefont {M.}~\bibnamefont {Dendzik}}, \bibinfo {author} {\bibfnamefont {J.~D.}\ \bibnamefont {Ziegler}}, \bibinfo {author} {\bibfnamefont {J.}~\bibnamefont {Maklar}}, \bibinfo {author} {\bibfnamefont {R.~P.}\ \bibnamefont {Xian}}, \bibinfo {author} {\bibfnamefont {A.}~\bibnamefont {Neef}}, \bibinfo {author} {\bibfnamefont {A.}~\bibnamefont {Mohammed}}, \bibinfo {author} {\bibfnamefont {A.}~\bibnamefont {Schulz}}, \bibinfo {author} {\bibfnamefont {M.}~\bibnamefont {Stadler}}, \bibinfo {author} {\bibfnamefont {M.}~\bibnamefont {Jetter}}, \bibinfo {author} {\bibfnamefont {P.}~\bibnamefont
  {Michler}}, \bibinfo {author} {\bibfnamefont {T.}~\bibnamefont {Taniguchi}}, \bibinfo {author} {\bibfnamefont {K.}~\bibnamefont {Watanabe}}, \bibinfo {author} {\bibfnamefont {H.}~\bibnamefont {Takagi}}, \bibinfo {author} {\bibfnamefont {U.}~\bibnamefont {Starke}}, \bibinfo {author} {\bibfnamefont {A.}~\bibnamefont {Chernikov}}, \bibinfo {author} {\bibfnamefont {M.}~\bibnamefont {Wolf}}, \bibinfo {author} {\bibfnamefont {H.}~\bibnamefont {Nakamura}}, \bibinfo {author} {\bibfnamefont {A.}~\bibnamefont {Knorr}}, \bibinfo {author} {\bibfnamefont {L.}~\bibnamefont {Rettig}},\ and\ \bibinfo {author} {\bibfnamefont {R.}~\bibnamefont {Ernstorfer}},\ }\bibfield  {title} {\bibinfo {title} {Observation of ultrafast interfacial {{Meitner-Auger}} energy transfer in a {{Van}} der {{Waals}} heterostructure},\ }\href {https://doi.org/10.1038/s41467-023-40815-8} {\bibfield  {journal} {\bibinfo  {journal} {Nature Communications}\ }\textbf {\bibinfo {volume} {14}},\ \bibinfo {pages} {5057} (\bibinfo {year}
  {2023})}\BibitemShut {NoStop}%
\bibitem [{\citenamefont {Weston}\ \emph {et~al.}(2020)\citenamefont {Weston}, \citenamefont {Zou}, \citenamefont {Enaldiev}, \citenamefont {Summerfield}, \citenamefont {Clark}, \citenamefont {Z{\'o}lyomi}, \citenamefont {Graham}, \citenamefont {Yelgel}, \citenamefont {Magorrian}, \citenamefont {Zhou}, \citenamefont {Zultak}, \citenamefont {Hopkinson}, \citenamefont {Barinov}, \citenamefont {Bointon}, \citenamefont {Kretinin}, \citenamefont {Wilson}, \citenamefont {Beton}, \citenamefont {Fal'ko}, \citenamefont {Haigh},\ and\ \citenamefont {Gorbachev}}]{weston_atomic_2020}%
  \BibitemOpen
  \bibfield  {author} {\bibinfo {author} {\bibfnamefont {A.}~\bibnamefont {Weston}}, \bibinfo {author} {\bibfnamefont {Y.}~\bibnamefont {Zou}}, \bibinfo {author} {\bibfnamefont {V.}~\bibnamefont {Enaldiev}}, \bibinfo {author} {\bibfnamefont {A.}~\bibnamefont {Summerfield}}, \bibinfo {author} {\bibfnamefont {N.}~\bibnamefont {Clark}}, \bibinfo {author} {\bibfnamefont {V.}~\bibnamefont {Z{\'o}lyomi}}, \bibinfo {author} {\bibfnamefont {A.}~\bibnamefont {Graham}}, \bibinfo {author} {\bibfnamefont {C.}~\bibnamefont {Yelgel}}, \bibinfo {author} {\bibfnamefont {S.}~\bibnamefont {Magorrian}}, \bibinfo {author} {\bibfnamefont {M.}~\bibnamefont {Zhou}}, \bibinfo {author} {\bibfnamefont {J.}~\bibnamefont {Zultak}}, \bibinfo {author} {\bibfnamefont {D.}~\bibnamefont {Hopkinson}}, \bibinfo {author} {\bibfnamefont {A.}~\bibnamefont {Barinov}}, \bibinfo {author} {\bibfnamefont {T.~H.}\ \bibnamefont {Bointon}}, \bibinfo {author} {\bibfnamefont {A.}~\bibnamefont {Kretinin}}, \bibinfo {author} {\bibfnamefont {N.~R.}\
  \bibnamefont {Wilson}}, \bibinfo {author} {\bibfnamefont {P.~H.}\ \bibnamefont {Beton}}, \bibinfo {author} {\bibfnamefont {V.~I.}\ \bibnamefont {Fal'ko}}, \bibinfo {author} {\bibfnamefont {S.~J.}\ \bibnamefont {Haigh}},\ and\ \bibinfo {author} {\bibfnamefont {R.}~\bibnamefont {Gorbachev}},\ }\bibfield  {title} {\bibinfo {title} {Atomic reconstruction in twisted bilayers of transition metal dichalcogenides},\ }\href {https://doi.org/10.1038/s41565-020-0682-9} {\bibfield  {journal} {\bibinfo  {journal} {Nature Nanotechnology}\ }\textbf {\bibinfo {volume} {15}},\ \bibinfo {pages} {592} (\bibinfo {year} {2020})}\BibitemShut {NoStop}%
\bibitem [{\citenamefont {Rosenberger}\ \emph {et~al.}(2020)\citenamefont {Rosenberger}, \citenamefont {Chuang}, \citenamefont {Phillips}, \citenamefont {Oleshko}, \citenamefont {McCreary}, \citenamefont {Sivaram}, \citenamefont {Hellberg},\ and\ \citenamefont {Jonker}}]{rosenberger_twist_2020}%
  \BibitemOpen
  \bibfield  {author} {\bibinfo {author} {\bibfnamefont {M.~R.}\ \bibnamefont {Rosenberger}}, \bibinfo {author} {\bibfnamefont {H.-J.}\ \bibnamefont {Chuang}}, \bibinfo {author} {\bibfnamefont {M.}~\bibnamefont {Phillips}}, \bibinfo {author} {\bibfnamefont {V.~P.}\ \bibnamefont {Oleshko}}, \bibinfo {author} {\bibfnamefont {K.~M.}\ \bibnamefont {McCreary}}, \bibinfo {author} {\bibfnamefont {S.~V.}\ \bibnamefont {Sivaram}}, \bibinfo {author} {\bibfnamefont {C.~S.}\ \bibnamefont {Hellberg}},\ and\ \bibinfo {author} {\bibfnamefont {B.~T.}\ \bibnamefont {Jonker}},\ }\bibfield  {title} {\bibinfo {title} {Twist {{Angle-Dependent Atomic Reconstruction}} and {{Moir{\'e} Patterns}} in {{Transition Metal Dichalcogenide Heterostructures}}},\ }\href {https://doi.org/10.1021/acsnano.0c00088} {\bibfield  {journal} {\bibinfo  {journal} {ACS Nano}\ }\textbf {\bibinfo {volume} {14}},\ \bibinfo {pages} {4550} (\bibinfo {year} {2020})}\BibitemShut {NoStop}%
\bibitem [{\citenamefont {Pierucci}\ \emph {et~al.}(2016)\citenamefont {Pierucci}, \citenamefont {Henck}, \citenamefont {Avila}, \citenamefont {Balan}, \citenamefont {Naylor}, \citenamefont {Patriarche}, \citenamefont {Dappe}, \citenamefont {Silly}, \citenamefont {Sirotti}, \citenamefont {Johnson}, \citenamefont {Asensio},\ and\ \citenamefont {Ouerghi}}]{pierucci_band_2016}%
  \BibitemOpen
  \bibfield  {author} {\bibinfo {author} {\bibfnamefont {D.}~\bibnamefont {Pierucci}}, \bibinfo {author} {\bibfnamefont {H.}~\bibnamefont {Henck}}, \bibinfo {author} {\bibfnamefont {J.}~\bibnamefont {Avila}}, \bibinfo {author} {\bibfnamefont {A.}~\bibnamefont {Balan}}, \bibinfo {author} {\bibfnamefont {C.~H.}\ \bibnamefont {Naylor}}, \bibinfo {author} {\bibfnamefont {G.}~\bibnamefont {Patriarche}}, \bibinfo {author} {\bibfnamefont {Y.~J.}\ \bibnamefont {Dappe}}, \bibinfo {author} {\bibfnamefont {M.~G.}\ \bibnamefont {Silly}}, \bibinfo {author} {\bibfnamefont {F.}~\bibnamefont {Sirotti}}, \bibinfo {author} {\bibfnamefont {A.~T.~C.}\ \bibnamefont {Johnson}}, \bibinfo {author} {\bibfnamefont {M.~C.}\ \bibnamefont {Asensio}},\ and\ \bibinfo {author} {\bibfnamefont {A.}~\bibnamefont {Ouerghi}},\ }\bibfield  {title} {\bibinfo {title} {Band {{Alignment}} and {{Minigaps}} in {{Monolayer MoS2-Graphene}} van der {{Waals Heterostructures}}},\ }\href {https://doi.org/10.1021/acs.nanolett.6b00609} {\bibfield  {journal}
  {\bibinfo  {journal} {Nano Letters}\ }\textbf {\bibinfo {volume} {16}},\ \bibinfo {pages} {4054} (\bibinfo {year} {2016})}\BibitemShut {NoStop}%
\bibitem [{\citenamefont {Wilson}\ \emph {et~al.}(2017)\citenamefont {Wilson}, \citenamefont {Nguyen}, \citenamefont {Seyler}, \citenamefont {Rivera}, \citenamefont {Marsden}, \citenamefont {Laker}, \citenamefont {Constantinescu}, \citenamefont {Kandyba}, \citenamefont {Barinov}, \citenamefont {Hine}, \citenamefont {Xu},\ and\ \citenamefont {Cobden}}]{wilson_determination_2017}%
  \BibitemOpen
  \bibfield  {author} {\bibinfo {author} {\bibfnamefont {N.~R.}\ \bibnamefont {Wilson}}, \bibinfo {author} {\bibfnamefont {P.~V.}\ \bibnamefont {Nguyen}}, \bibinfo {author} {\bibfnamefont {K.}~\bibnamefont {Seyler}}, \bibinfo {author} {\bibfnamefont {P.}~\bibnamefont {Rivera}}, \bibinfo {author} {\bibfnamefont {A.~J.}\ \bibnamefont {Marsden}}, \bibinfo {author} {\bibfnamefont {Z.~P.~L.}\ \bibnamefont {Laker}}, \bibinfo {author} {\bibfnamefont {G.~C.}\ \bibnamefont {Constantinescu}}, \bibinfo {author} {\bibfnamefont {V.}~\bibnamefont {Kandyba}}, \bibinfo {author} {\bibfnamefont {A.}~\bibnamefont {Barinov}}, \bibinfo {author} {\bibfnamefont {N.~D.~M.}\ \bibnamefont {Hine}}, \bibinfo {author} {\bibfnamefont {X.}~\bibnamefont {Xu}},\ and\ \bibinfo {author} {\bibfnamefont {D.~H.}\ \bibnamefont {Cobden}},\ }\bibfield  {title} {\bibinfo {title} {Determination of band offsets, hybridization, and exciton binding in {{2D}} semiconductor heterostructures},\ }\href {https://doi.org/10.1126/sciadv.1601832} {\bibfield
  {journal} {\bibinfo  {journal} {Science Advances}\ }\textbf {\bibinfo {volume} {3}},\ \bibinfo {pages} {e1601832} (\bibinfo {year} {2017})}\BibitemShut {NoStop}%
\bibitem [{\citenamefont {Henck}\ \emph {et~al.}(2018)\citenamefont {Henck}, \citenamefont {Ben~Aziza}, \citenamefont {Pierucci}, \citenamefont {Laourine}, \citenamefont {Reale}, \citenamefont {Palczynski}, \citenamefont {Chaste}, \citenamefont {Silly}, \citenamefont {Bertran}, \citenamefont {Le~F{\`e}vre}, \citenamefont {Lhuillier}, \citenamefont {Wakamura}, \citenamefont {Mattevi}, \citenamefont {Rault}, \citenamefont {Calandra},\ and\ \citenamefont {Ouerghi}}]{henck_electronic_2018}%
  \BibitemOpen
  \bibfield  {author} {\bibinfo {author} {\bibfnamefont {H.}~\bibnamefont {Henck}}, \bibinfo {author} {\bibfnamefont {Z.}~\bibnamefont {Ben~Aziza}}, \bibinfo {author} {\bibfnamefont {D.}~\bibnamefont {Pierucci}}, \bibinfo {author} {\bibfnamefont {F.}~\bibnamefont {Laourine}}, \bibinfo {author} {\bibfnamefont {F.}~\bibnamefont {Reale}}, \bibinfo {author} {\bibfnamefont {P.}~\bibnamefont {Palczynski}}, \bibinfo {author} {\bibfnamefont {J.}~\bibnamefont {Chaste}}, \bibinfo {author} {\bibfnamefont {M.~G.}\ \bibnamefont {Silly}}, \bibinfo {author} {\bibfnamefont {F.}~\bibnamefont {Bertran}}, \bibinfo {author} {\bibfnamefont {P.}~\bibnamefont {Le~F{\`e}vre}}, \bibinfo {author} {\bibfnamefont {E.}~\bibnamefont {Lhuillier}}, \bibinfo {author} {\bibfnamefont {T.}~\bibnamefont {Wakamura}}, \bibinfo {author} {\bibfnamefont {C.}~\bibnamefont {Mattevi}}, \bibinfo {author} {\bibfnamefont {J.~E.}\ \bibnamefont {Rault}}, \bibinfo {author} {\bibfnamefont {M.}~\bibnamefont {Calandra}},\ and\ \bibinfo {author} {\bibfnamefont
  {A.}~\bibnamefont {Ouerghi}},\ }\bibfield  {title} {\bibinfo {title} {Electronic band structure of {Two-Dimensional} {WS2}/graphene van der {{Waals Heterostructures}}},\ }\href {https://doi.org/10.1103/PhysRevB.97.155421} {\bibfield  {journal} {\bibinfo  {journal} {Physical Review B}\ }\textbf {\bibinfo {volume} {97}},\ \bibinfo {pages} {155421} (\bibinfo {year} {2018})}\BibitemShut {NoStop}%
\bibitem [{\citenamefont {Scalise}\ \emph {et~al.}(2012)\citenamefont {Scalise}, \citenamefont {Houssa}, \citenamefont {Pourtois}, \citenamefont {Afanas'ev},\ and\ \citenamefont {Stesmans}}]{scalise_strain-induced_2012}%
  \BibitemOpen
  \bibfield  {author} {\bibinfo {author} {\bibfnamefont {E.}~\bibnamefont {Scalise}}, \bibinfo {author} {\bibfnamefont {M.}~\bibnamefont {Houssa}}, \bibinfo {author} {\bibfnamefont {G.}~\bibnamefont {Pourtois}}, \bibinfo {author} {\bibfnamefont {V.}~\bibnamefont {Afanas'ev}},\ and\ \bibinfo {author} {\bibfnamefont {A.}~\bibnamefont {Stesmans}},\ }\bibfield  {title} {\bibinfo {title} {Strain-induced semiconductor to metal transition in the two-dimensional honeycomb structure of {{MoS2}}},\ }\href {https://doi.org/10.1007/s12274-011-0183-0} {\bibfield  {journal} {\bibinfo  {journal} {Nano Research}\ }\textbf {\bibinfo {volume} {5}},\ \bibinfo {pages} {43} (\bibinfo {year} {2012})}\BibitemShut {NoStop}%
\bibitem [{\citenamefont {Yun}\ \emph {et~al.}(2012)\citenamefont {Yun}, \citenamefont {Han}, \citenamefont {Hong}, \citenamefont {Kim},\ and\ \citenamefont {Lee}}]{yun_thickness_2012}%
  \BibitemOpen
  \bibfield  {author} {\bibinfo {author} {\bibfnamefont {W.~S.}\ \bibnamefont {Yun}}, \bibinfo {author} {\bibfnamefont {S.~W.}\ \bibnamefont {Han}}, \bibinfo {author} {\bibfnamefont {S.~C.}\ \bibnamefont {Hong}}, \bibinfo {author} {\bibfnamefont {I.~G.}\ \bibnamefont {Kim}},\ and\ \bibinfo {author} {\bibfnamefont {J.~D.}\ \bibnamefont {Lee}},\ }\bibfield  {title} {\bibinfo {title} {Thickness and strain effects on electronic structures of transition metal dichalcogenides: {2H-}{M}x2 semiconductors ({M}={Mo}, {{W}}; {X}= {{S}}, {{Se}}, {{Te}})},\ }\href {https://doi.org/10.1103/PhysRevB.85.033305} {\bibfield  {journal} {\bibinfo  {journal} {Physical Review B}\ }\textbf {\bibinfo {volume} {85}},\ \bibinfo {pages} {033305} (\bibinfo {year} {2012})}\BibitemShut {NoStop}%
\bibitem [{\citenamefont {Ebnonnasir}\ \emph {et~al.}(2014)\citenamefont {Ebnonnasir}, \citenamefont {Narayanan}, \citenamefont {Kodambaka},\ and\ \citenamefont {Ciobanu}}]{ebnonnasir_tunable_2014}%
  \BibitemOpen
  \bibfield  {author} {\bibinfo {author} {\bibfnamefont {A.}~\bibnamefont {Ebnonnasir}}, \bibinfo {author} {\bibfnamefont {B.}~\bibnamefont {Narayanan}}, \bibinfo {author} {\bibfnamefont {S.}~\bibnamefont {Kodambaka}},\ and\ \bibinfo {author} {\bibfnamefont {C.~V.}\ \bibnamefont {Ciobanu}},\ }\bibfield  {title} {\bibinfo {title} {Tunable {{MoS2}} bandgap in {{MoS2-graphene}} heterostructures},\ }\href {https://doi.org/10.1063/1.4891430} {\bibfield  {journal} {\bibinfo  {journal} {Applied Physics Letters}\ }\textbf {\bibinfo {volume} {105}},\ \bibinfo {pages} {031603} (\bibinfo {year} {2014})}\BibitemShut {NoStop}%
\bibitem [{\citenamefont {Junior}\ \emph {et~al.}(2023)\citenamefont {Junior}, \citenamefont {Naimer}, \citenamefont {McCreary}, \citenamefont {Jonker}, \citenamefont {Finley}, \citenamefont {Crooker}, \citenamefont {Fabian},\ and\ \citenamefont {Stier}}]{junior_proximity-enhanced_2023}%
  \BibitemOpen
  \bibfield  {author} {\bibinfo {author} {\bibfnamefont {P.~E.~F.}\ \bibnamefont {Junior}}, \bibinfo {author} {\bibfnamefont {T.}~\bibnamefont {Naimer}}, \bibinfo {author} {\bibfnamefont {K.~M.}\ \bibnamefont {McCreary}}, \bibinfo {author} {\bibfnamefont {B.~T.}\ \bibnamefont {Jonker}}, \bibinfo {author} {\bibfnamefont {J.~J.}\ \bibnamefont {Finley}}, \bibinfo {author} {\bibfnamefont {S.~A.}\ \bibnamefont {Crooker}}, \bibinfo {author} {\bibfnamefont {J.}~\bibnamefont {Fabian}},\ and\ \bibinfo {author} {\bibfnamefont {A.~V.}\ \bibnamefont {Stier}},\ }\bibfield  {title} {\bibinfo {title} {Proximity-enhanced valley {{Zeeman}} splitting at the {{WS2}}/graphene interface},\ }\href {https://doi.org/10.1088/2053-1583/acd5df} {\bibfield  {journal} {\bibinfo  {journal} {2D Materials}\ }\textbf {\bibinfo {volume} {10}},\ \bibinfo {pages} {034002} (\bibinfo {year} {2023})}\BibitemShut {NoStop}%
\bibitem [{\citenamefont {Magnozzi}\ \emph {et~al.}(2020)\citenamefont {Magnozzi}, \citenamefont {Ferrera}, \citenamefont {Piccinini}, \citenamefont {Pace}, \citenamefont {Forti}, \citenamefont {Fabbri}, \citenamefont {Coletti}, \citenamefont {Bisio},\ and\ \citenamefont {Canepa}}]{magnozzi_optical_2020}%
  \BibitemOpen
  \bibfield  {author} {\bibinfo {author} {\bibfnamefont {M.}~\bibnamefont {Magnozzi}}, \bibinfo {author} {\bibfnamefont {M.}~\bibnamefont {Ferrera}}, \bibinfo {author} {\bibfnamefont {G.}~\bibnamefont {Piccinini}}, \bibinfo {author} {\bibfnamefont {S.}~\bibnamefont {Pace}}, \bibinfo {author} {\bibfnamefont {S.}~\bibnamefont {Forti}}, \bibinfo {author} {\bibfnamefont {F.}~\bibnamefont {Fabbri}}, \bibinfo {author} {\bibfnamefont {C.}~\bibnamefont {Coletti}}, \bibinfo {author} {\bibfnamefont {F.}~\bibnamefont {Bisio}},\ and\ \bibinfo {author} {\bibfnamefont {M.}~\bibnamefont {Canepa}},\ }\bibfield  {title} {\bibinfo {title} {Optical dielectric function of two-dimensional {{WS}}2 on epitaxial graphene},\ }\href {https://doi.org/10.1088/2053-1583/ab6f0b} {\bibfield  {journal} {\bibinfo  {journal} {2D Materials}\ }\textbf {\bibinfo {volume} {7}},\ \bibinfo {pages} {025024} (\bibinfo {year} {2020})}\BibitemShut {NoStop}%
\bibitem [{\citenamefont {{Riis-Jensen}}\ \emph {et~al.}(2020)\citenamefont {{Riis-Jensen}}, \citenamefont {Lu},\ and\ \citenamefont {Thygesen}}]{riis-jensen_electrically_2020}%
  \BibitemOpen
  \bibfield  {author} {\bibinfo {author} {\bibfnamefont {A.~C.}\ \bibnamefont {{Riis-Jensen}}}, \bibinfo {author} {\bibfnamefont {J.}~\bibnamefont {Lu}},\ and\ \bibinfo {author} {\bibfnamefont {K.~S.}\ \bibnamefont {Thygesen}},\ }\bibfield  {title} {\bibinfo {title} {Electrically controlled dielectric band gap engineering in a two-dimensional semiconductor},\ }\href {https://doi.org/10.1103/PhysRevB.101.121110} {\bibfield  {journal} {\bibinfo  {journal} {Physical Review B}\ }\textbf {\bibinfo {volume} {101}},\ \bibinfo {pages} {121110} (\bibinfo {year} {2020})}\BibitemShut {NoStop}%
\bibitem [{\citenamefont {Tebbe}\ \emph {et~al.}(2023)\citenamefont {Tebbe}, \citenamefont {Sch{\"u}tte}, \citenamefont {Watanabe}, \citenamefont {Taniguchi}, \citenamefont {Stampfer}, \citenamefont {Beschoten},\ and\ \citenamefont {Waldecker}}]{tebbe_tailoring_2023}%
  \BibitemOpen
  \bibfield  {author} {\bibinfo {author} {\bibfnamefont {D.}~\bibnamefont {Tebbe}}, \bibinfo {author} {\bibfnamefont {M.}~\bibnamefont {Sch{\"u}tte}}, \bibinfo {author} {\bibfnamefont {K.}~\bibnamefont {Watanabe}}, \bibinfo {author} {\bibfnamefont {T.}~\bibnamefont {Taniguchi}}, \bibinfo {author} {\bibfnamefont {C.}~\bibnamefont {Stampfer}}, \bibinfo {author} {\bibfnamefont {B.}~\bibnamefont {Beschoten}},\ and\ \bibinfo {author} {\bibfnamefont {L.}~\bibnamefont {Waldecker}},\ }\bibfield  {title} {\bibinfo {title} {Tailoring the dielectric screening in {{WS2}}{\textendash}graphene heterostructures},\ }\href {https://doi.org/10.1038/s41699-023-00394-0} {\bibfield  {journal} {\bibinfo  {journal} {npj 2D Materials and Applications}\ }\textbf {\bibinfo {volume} {7}},\ \bibinfo {pages} {1} (\bibinfo {year} {2023})}\BibitemShut {NoStop}%
\bibitem [{\citenamefont {Latini}\ \emph {et~al.}(2015)\citenamefont {Latini}, \citenamefont {Olsen},\ and\ \citenamefont {Thygesen}}]{latini_excitons_2015}%
  \BibitemOpen
  \bibfield  {author} {\bibinfo {author} {\bibfnamefont {S.}~\bibnamefont {Latini}}, \bibinfo {author} {\bibfnamefont {T.}~\bibnamefont {Olsen}},\ and\ \bibinfo {author} {\bibfnamefont {K.~S.}\ \bibnamefont {Thygesen}},\ }\bibfield  {title} {\bibinfo {title} {Excitons in van der {{Waals}} heterostructures: {{The}} important role of dielectric screening},\ }\href {https://doi.org/10.1103/PhysRevB.92.245123} {\bibfield  {journal} {\bibinfo  {journal} {Physical Review B}\ }\textbf {\bibinfo {volume} {92}},\ \bibinfo {pages} {245123} (\bibinfo {year} {2015})}\BibitemShut {NoStop}%
\bibitem [{\citenamefont {Karni}\ \emph {et~al.}(2022)\citenamefont {Karni}, \citenamefont {Barr{\'e}}, \citenamefont {Pareek}, \citenamefont {Georgaras}, \citenamefont {Man}, \citenamefont {Sahoo}, \citenamefont {Bacon}, \citenamefont {Zhu}, \citenamefont {Ribeiro}, \citenamefont {O’Beirne} \emph {et~al.}}]{karni2022structure}%
  \BibitemOpen
  \bibfield  {author} {\bibinfo {author} {\bibfnamefont {O.}~\bibnamefont {Karni}}, \bibinfo {author} {\bibfnamefont {E.}~\bibnamefont {Barr{\'e}}}, \bibinfo {author} {\bibfnamefont {V.}~\bibnamefont {Pareek}}, \bibinfo {author} {\bibfnamefont {J.~D.}\ \bibnamefont {Georgaras}}, \bibinfo {author} {\bibfnamefont {M.~K.}\ \bibnamefont {Man}}, \bibinfo {author} {\bibfnamefont {C.}~\bibnamefont {Sahoo}}, \bibinfo {author} {\bibfnamefont {D.~R.}\ \bibnamefont {Bacon}}, \bibinfo {author} {\bibfnamefont {X.}~\bibnamefont {Zhu}}, \bibinfo {author} {\bibfnamefont {H.~B.}\ \bibnamefont {Ribeiro}}, \bibinfo {author} {\bibfnamefont {A.~L.}\ \bibnamefont {O’Beirne}}, \emph {et~al.},\ }\bibfield  {title} {\bibinfo {title} {Structure of the moir{\'e} exciton captured by imaging its electron and hole},\ }\href@noop {} {\bibfield  {journal} {\bibinfo  {journal} {Nature}\ }\textbf {\bibinfo {volume} {603}},\ \bibinfo {pages} {247} (\bibinfo {year} {2022})}\BibitemShut {NoStop}%
\bibitem [{\citenamefont {Naik}\ \emph {et~al.}(2022)\citenamefont {Naik}, \citenamefont {Regan}, \citenamefont {Zhang}, \citenamefont {Chan}, \citenamefont {Li}, \citenamefont {Wang}, \citenamefont {Yoon}, \citenamefont {Ong}, \citenamefont {Zhao}, \citenamefont {Zhao}, \citenamefont {Utama}, \citenamefont {Gao}, \citenamefont {Wei}, \citenamefont {Sayyad}, \citenamefont {Yumigeta}, \citenamefont {Watanabe}, \citenamefont {Taniguchi}, \citenamefont {Tongay}, \citenamefont {{da Jornada}}, \citenamefont {Wang},\ and\ \citenamefont {Louie}}]{naik_intralayer_2022}%
  \BibitemOpen
  \bibfield  {author} {\bibinfo {author} {\bibfnamefont {M.~H.}\ \bibnamefont {Naik}}, \bibinfo {author} {\bibfnamefont {E.~C.}\ \bibnamefont {Regan}}, \bibinfo {author} {\bibfnamefont {Z.}~\bibnamefont {Zhang}}, \bibinfo {author} {\bibfnamefont {Y.-H.}\ \bibnamefont {Chan}}, \bibinfo {author} {\bibfnamefont {Z.}~\bibnamefont {Li}}, \bibinfo {author} {\bibfnamefont {D.}~\bibnamefont {Wang}}, \bibinfo {author} {\bibfnamefont {Y.}~\bibnamefont {Yoon}}, \bibinfo {author} {\bibfnamefont {C.~S.}\ \bibnamefont {Ong}}, \bibinfo {author} {\bibfnamefont {W.}~\bibnamefont {Zhao}}, \bibinfo {author} {\bibfnamefont {S.}~\bibnamefont {Zhao}}, \bibinfo {author} {\bibfnamefont {M.~I.~B.}\ \bibnamefont {Utama}}, \bibinfo {author} {\bibfnamefont {B.}~\bibnamefont {Gao}}, \bibinfo {author} {\bibfnamefont {X.}~\bibnamefont {Wei}}, \bibinfo {author} {\bibfnamefont {M.}~\bibnamefont {Sayyad}}, \bibinfo {author} {\bibfnamefont {K.}~\bibnamefont {Yumigeta}}, \bibinfo {author} {\bibfnamefont {K.}~\bibnamefont {Watanabe}}, \bibinfo
  {author} {\bibfnamefont {T.}~\bibnamefont {Taniguchi}}, \bibinfo {author} {\bibfnamefont {S.}~\bibnamefont {Tongay}}, \bibinfo {author} {\bibfnamefont {F.~H.}\ \bibnamefont {{da Jornada}}}, \bibinfo {author} {\bibfnamefont {F.}~\bibnamefont {Wang}},\ and\ \bibinfo {author} {\bibfnamefont {S.~G.}\ \bibnamefont {Louie}},\ }\bibfield  {title} {\bibinfo {title} {Intralayer charge-transfer moir{\'e} excitons in van der {{Waals}} superlattices},\ }\href {https://doi.org/10.1038/s41586-022-04991-9} {\bibfield  {journal} {\bibinfo  {journal} {Nature}\ }\textbf {\bibinfo {volume} {609}},\ \bibinfo {pages} {52} (\bibinfo {year} {2022})}\BibitemShut {NoStop}%
\bibitem [{\citenamefont {Barr{\'e}}\ \emph {et~al.}(2022)\citenamefont {Barr{\'e}}, \citenamefont {Karni}, \citenamefont {Liu}, \citenamefont {O'Beirne}, \citenamefont {Chen}, \citenamefont {Ribeiro}, \citenamefont {Yu}, \citenamefont {Kim}, \citenamefont {Watanabe}, \citenamefont {Taniguchi}, \citenamefont {Barmak}, \citenamefont {Lui}, \citenamefont {{Refaely-Abramson}}, \citenamefont {{da Jornada}},\ and\ \citenamefont {Heinz}}]{barre_optical_2022}%
  \BibitemOpen
  \bibfield  {author} {\bibinfo {author} {\bibfnamefont {E.}~\bibnamefont {Barr{\'e}}}, \bibinfo {author} {\bibfnamefont {O.}~\bibnamefont {Karni}}, \bibinfo {author} {\bibfnamefont {E.}~\bibnamefont {Liu}}, \bibinfo {author} {\bibfnamefont {A.~L.}\ \bibnamefont {O'Beirne}}, \bibinfo {author} {\bibfnamefont {X.}~\bibnamefont {Chen}}, \bibinfo {author} {\bibfnamefont {H.~B.}\ \bibnamefont {Ribeiro}}, \bibinfo {author} {\bibfnamefont {L.}~\bibnamefont {Yu}}, \bibinfo {author} {\bibfnamefont {B.}~\bibnamefont {Kim}}, \bibinfo {author} {\bibfnamefont {K.}~\bibnamefont {Watanabe}}, \bibinfo {author} {\bibfnamefont {T.}~\bibnamefont {Taniguchi}}, \bibinfo {author} {\bibfnamefont {K.}~\bibnamefont {Barmak}}, \bibinfo {author} {\bibfnamefont {C.~H.}\ \bibnamefont {Lui}}, \bibinfo {author} {\bibfnamefont {S.}~\bibnamefont {{Refaely-Abramson}}}, \bibinfo {author} {\bibfnamefont {F.~H.}\ \bibnamefont {{da Jornada}}},\ and\ \bibinfo {author} {\bibfnamefont {T.~F.}\ \bibnamefont {Heinz}},\ }\bibfield  {title} {\bibinfo
  {title} {Optical absorption of interlayer excitons in transition-metal dichalcogenide heterostructures},\ }\href {https://doi.org/10.1126/science.abm8511} {\bibfield  {journal} {\bibinfo  {journal} {Science}\ }\textbf {\bibinfo {volume} {376}},\ \bibinfo {pages} {406} (\bibinfo {year} {2022})}\BibitemShut {NoStop}%
\bibitem [{\citenamefont {Kundu}\ \emph {et~al.}(2023)\citenamefont {Kundu}, \citenamefont {Amit}, \citenamefont {Krishnamurthy}, \citenamefont {Jain},\ and\ \citenamefont {Refaely-Abramson}}]{kundu2023exciton}%
  \BibitemOpen
  \bibfield  {author} {\bibinfo {author} {\bibfnamefont {S.}~\bibnamefont {Kundu}}, \bibinfo {author} {\bibfnamefont {T.}~\bibnamefont {Amit}}, \bibinfo {author} {\bibfnamefont {H.}~\bibnamefont {Krishnamurthy}}, \bibinfo {author} {\bibfnamefont {M.}~\bibnamefont {Jain}},\ and\ \bibinfo {author} {\bibfnamefont {S.}~\bibnamefont {Refaely-Abramson}},\ }\bibfield  {title} {\bibinfo {title} {Exciton fine structure in twisted transition metal dichalcogenide heterostructures},\ }\href@noop {} {\bibfield  {journal} {\bibinfo  {journal} {npj Computational Materials}\ }\textbf {\bibinfo {volume} {9}},\ \bibinfo {pages} {186} (\bibinfo {year} {2023})}\BibitemShut {NoStop}%
\bibitem [{\citenamefont {Kundu}\ \emph {et~al.}(2022)\citenamefont {Kundu}, \citenamefont {Naik}, \citenamefont {Krishnamurthy},\ and\ \citenamefont {Jain}}]{kundu2022moire}%
  \BibitemOpen
  \bibfield  {author} {\bibinfo {author} {\bibfnamefont {S.}~\bibnamefont {Kundu}}, \bibinfo {author} {\bibfnamefont {M.~H.}\ \bibnamefont {Naik}}, \bibinfo {author} {\bibfnamefont {H.}~\bibnamefont {Krishnamurthy}},\ and\ \bibinfo {author} {\bibfnamefont {M.}~\bibnamefont {Jain}},\ }\bibfield  {title} {\bibinfo {title} {Moir{\'e} induced topology and flat bands in twisted bilayer wse 2: A first-principles study},\ }\href@noop {} {\bibfield  {journal} {\bibinfo  {journal} {Physical Review B}\ }\textbf {\bibinfo {volume} {105}},\ \bibinfo {pages} {L081108} (\bibinfo {year} {2022})}\BibitemShut {NoStop}%
\bibitem [{\citenamefont {{Hernang{\'o}mez-P{\'e}rez}}\ \emph {et~al.}(2023)\citenamefont {{Hernang{\'o}mez-P{\'e}rez}}, \citenamefont {Kleiner},\ and\ \citenamefont {{Refaely-Abramson}}}]{hernangomez-perez_reduced_2023}%
  \BibitemOpen
  \bibfield  {author} {\bibinfo {author} {\bibfnamefont {D.}~\bibnamefont {{Hernang{\'o}mez-P{\'e}rez}}}, \bibinfo {author} {\bibfnamefont {A.}~\bibnamefont {Kleiner}},\ and\ \bibinfo {author} {\bibfnamefont {S.}~\bibnamefont {{Refaely-Abramson}}},\ }\bibfield  {title} {\bibinfo {title} {Reduced {{Absorption Due}} to {{Defect-Localized Interlayer Excitons}} in {{Transition-Metal Dichalcogenide}}{\textendash}{{Graphene Heterostructures}}},\ }\href {https://doi.org/10.1021/acs.nanolett.3c01182} {\bibfield  {journal} {\bibinfo  {journal} {Nano Letters}\ }\textbf {\bibinfo {volume} {23}},\ \bibinfo {pages} {5995} (\bibinfo {year} {2023})}\BibitemShut {NoStop}%
\bibitem [{\citenamefont {{Refaely-Abramson}}\ \emph {et~al.}(2018)\citenamefont {{Refaely-Abramson}}, \citenamefont {Qiu}, \citenamefont {Louie},\ and\ \citenamefont {Neaton}}]{refaely-abramson_defect-induced_2018}%
  \BibitemOpen
  \bibfield  {author} {\bibinfo {author} {\bibfnamefont {S.}~\bibnamefont {{Refaely-Abramson}}}, \bibinfo {author} {\bibfnamefont {D.~Y.}\ \bibnamefont {Qiu}}, \bibinfo {author} {\bibfnamefont {S.~G.}\ \bibnamefont {Louie}},\ and\ \bibinfo {author} {\bibfnamefont {J.~B.}\ \bibnamefont {Neaton}},\ }\bibfield  {title} {\bibinfo {title} {Defect-{{Induced Modification}} of {{Low-Lying Excitons}} and {{Valley Selectivity}} in {{Monolayer Transition Metal Dichalcogenides}}},\ }\href {https://doi.org/10.1103/PhysRevLett.121.167402} {\bibfield  {journal} {\bibinfo  {journal} {Physical Review Letters}\ }\textbf {\bibinfo {volume} {121}},\ \bibinfo {pages} {167402} (\bibinfo {year} {2018})}\BibitemShut {NoStop}%
\bibitem [{\citenamefont {Amit}\ \emph {et~al.}(2022)\citenamefont {Amit}, \citenamefont {{Hernang{\'o}mez-P{\'e}rez}}, \citenamefont {Cohen}, \citenamefont {Qiu},\ and\ \citenamefont {{Refaely-Abramson}}}]{amit_tunable_2022}%
  \BibitemOpen
  \bibfield  {author} {\bibinfo {author} {\bibfnamefont {T.}~\bibnamefont {Amit}}, \bibinfo {author} {\bibfnamefont {D.}~\bibnamefont {{Hernang{\'o}mez-P{\'e}rez}}}, \bibinfo {author} {\bibfnamefont {G.}~\bibnamefont {Cohen}}, \bibinfo {author} {\bibfnamefont {D.~Y.}\ \bibnamefont {Qiu}},\ and\ \bibinfo {author} {\bibfnamefont {S.}~\bibnamefont {{Refaely-Abramson}}},\ }\bibfield  {title} {\bibinfo {title} {Tunable magneto-optical properties in mos$_2$ via defect-induced exciton transitions},\ }\href {https://doi.org/10.1103/PhysRevB.106.L161407} {\bibfield  {journal} {\bibinfo  {journal} {Physical Review B}\ }\textbf {\bibinfo {volume} {106}},\ \bibinfo {pages} {L161407} (\bibinfo {year} {2022})}\BibitemShut {NoStop}%
\bibitem [{\citenamefont {Kohn}\ and\ \citenamefont {Sham}(1965)}]{kohn_self-consistent_1965}%
  \BibitemOpen
  \bibfield  {author} {\bibinfo {author} {\bibfnamefont {W.}~\bibnamefont {Kohn}}\ and\ \bibinfo {author} {\bibfnamefont {L.~J.}\ \bibnamefont {Sham}},\ }\bibfield  {title} {\bibinfo {title} {Self-{{Consistent Equations Including Exchange}} and {{Correlation Effects}}},\ }\href {https://doi.org/10.1103/PhysRev.140.A1133} {\bibfield  {journal} {\bibinfo  {journal} {Physical Review}\ }\textbf {\bibinfo {volume} {140}},\ \bibinfo {pages} {A1133} (\bibinfo {year} {1965})}\BibitemShut {NoStop}%
\bibitem [{\citenamefont {Hybertsen}\ and\ \citenamefont {Louie}(1986)}]{hybertsen_electron_1986}%
  \BibitemOpen
  \bibfield  {author} {\bibinfo {author} {\bibfnamefont {M.~S.}\ \bibnamefont {Hybertsen}}\ and\ \bibinfo {author} {\bibfnamefont {S.~G.}\ \bibnamefont {Louie}},\ }\bibfield  {title} {\bibinfo {title} {Electron correlation in semiconductors and insulators: {{Band}} gaps and quasiparticle energies},\ }\href {https://doi.org/10.1103/PhysRevB.34.5390} {\bibfield  {journal} {\bibinfo  {journal} {Physical Review B}\ }\textbf {\bibinfo {volume} {34}},\ \bibinfo {pages} {5390} (\bibinfo {year} {1986})}\BibitemShut {NoStop}%
\bibitem [{\citenamefont {Raja}\ \emph {et~al.}(2017)\citenamefont {Raja}, \citenamefont {Chaves}, \citenamefont {Yu}, \citenamefont {Arefe}, \citenamefont {Hill}, \citenamefont {Rigosi}, \citenamefont {Berkelbach}, \citenamefont {Nagler}, \citenamefont {Sch{\"u}ller}, \citenamefont {Korn}, \citenamefont {Nuckolls}, \citenamefont {Hone}, \citenamefont {Brus}, \citenamefont {Heinz}, \citenamefont {Reichman},\ and\ \citenamefont {Chernikov}}]{raja_coulomb_2017}%
  \BibitemOpen
  \bibfield  {author} {\bibinfo {author} {\bibfnamefont {A.}~\bibnamefont {Raja}}, \bibinfo {author} {\bibfnamefont {A.}~\bibnamefont {Chaves}}, \bibinfo {author} {\bibfnamefont {J.}~\bibnamefont {Yu}}, \bibinfo {author} {\bibfnamefont {G.}~\bibnamefont {Arefe}}, \bibinfo {author} {\bibfnamefont {H.~M.}\ \bibnamefont {Hill}}, \bibinfo {author} {\bibfnamefont {A.~F.}\ \bibnamefont {Rigosi}}, \bibinfo {author} {\bibfnamefont {T.~C.}\ \bibnamefont {Berkelbach}}, \bibinfo {author} {\bibfnamefont {P.}~\bibnamefont {Nagler}}, \bibinfo {author} {\bibfnamefont {C.}~\bibnamefont {Sch{\"u}ller}}, \bibinfo {author} {\bibfnamefont {T.}~\bibnamefont {Korn}}, \bibinfo {author} {\bibfnamefont {C.}~\bibnamefont {Nuckolls}}, \bibinfo {author} {\bibfnamefont {J.}~\bibnamefont {Hone}}, \bibinfo {author} {\bibfnamefont {L.~E.}\ \bibnamefont {Brus}}, \bibinfo {author} {\bibfnamefont {T.~F.}\ \bibnamefont {Heinz}}, \bibinfo {author} {\bibfnamefont {D.~R.}\ \bibnamefont {Reichman}},\ and\ \bibinfo {author} {\bibfnamefont
  {A.}~\bibnamefont {Chernikov}},\ }\bibfield  {title} {\bibinfo {title} {Coulomb engineering of the bandgap and excitons in two-dimensional materials},\ }\href {https://doi.org/10.1038/ncomms15251} {\bibfield  {journal} {\bibinfo  {journal} {Nature Communications}\ }\textbf {\bibinfo {volume} {8}},\ \bibinfo {pages} {15251} (\bibinfo {year} {2017})}\BibitemShut {NoStop}%
\bibitem [{\citenamefont {Waldecker}\ \emph {et~al.}(2019)\citenamefont {Waldecker}, \citenamefont {Raja}, \citenamefont {R{\"o}sner}, \citenamefont {Steinke}, \citenamefont {Bostwick}, \citenamefont {Koch}, \citenamefont {Jozwiak}, \citenamefont {Taniguchi}, \citenamefont {Watanabe}, \citenamefont {Rotenberg}, \citenamefont {Wehling},\ and\ \citenamefont {Heinz}}]{waldecker_rigid_2019}%
  \BibitemOpen
  \bibfield  {author} {\bibinfo {author} {\bibfnamefont {L.}~\bibnamefont {Waldecker}}, \bibinfo {author} {\bibfnamefont {A.}~\bibnamefont {Raja}}, \bibinfo {author} {\bibfnamefont {M.}~\bibnamefont {R{\"o}sner}}, \bibinfo {author} {\bibfnamefont {C.}~\bibnamefont {Steinke}}, \bibinfo {author} {\bibfnamefont {A.}~\bibnamefont {Bostwick}}, \bibinfo {author} {\bibfnamefont {R.~J.}\ \bibnamefont {Koch}}, \bibinfo {author} {\bibfnamefont {C.}~\bibnamefont {Jozwiak}}, \bibinfo {author} {\bibfnamefont {T.}~\bibnamefont {Taniguchi}}, \bibinfo {author} {\bibfnamefont {K.}~\bibnamefont {Watanabe}}, \bibinfo {author} {\bibfnamefont {E.}~\bibnamefont {Rotenberg}}, \bibinfo {author} {\bibfnamefont {T.~O.}\ \bibnamefont {Wehling}},\ and\ \bibinfo {author} {\bibfnamefont {T.~F.}\ \bibnamefont {Heinz}},\ }\bibfield  {title} {\bibinfo {title} {Rigid {{Band Shifts}} in {{Two-Dimensional Semiconductors}} through {{External Dielectric Screening}}},\ }\href {https://doi.org/10.1103/PhysRevLett.123.206403} {\bibfield  {journal}
  {\bibinfo  {journal} {Physical Review Letters}\ }\textbf {\bibinfo {volume} {123}},\ \bibinfo {pages} {206403} (\bibinfo {year} {2019})}\BibitemShut {NoStop}%
\bibitem [{\citenamefont {Yuan}\ \emph {et~al.}(2018)\citenamefont {Yuan}, \citenamefont {Chung}, \citenamefont {Kuc}, \citenamefont {Wan}, \citenamefont {Xu}, \citenamefont {Chen}, \citenamefont {Heine},\ and\ \citenamefont {Huang}}]{yuan_photocarrier_2018}%
  \BibitemOpen
  \bibfield  {author} {\bibinfo {author} {\bibfnamefont {L.}~\bibnamefont {Yuan}}, \bibinfo {author} {\bibfnamefont {T.-F.}\ \bibnamefont {Chung}}, \bibinfo {author} {\bibfnamefont {A.}~\bibnamefont {Kuc}}, \bibinfo {author} {\bibfnamefont {Y.}~\bibnamefont {Wan}}, \bibinfo {author} {\bibfnamefont {Y.}~\bibnamefont {Xu}}, \bibinfo {author} {\bibfnamefont {Y.~P.}\ \bibnamefont {Chen}}, \bibinfo {author} {\bibfnamefont {T.}~\bibnamefont {Heine}},\ and\ \bibinfo {author} {\bibfnamefont {L.}~\bibnamefont {Huang}},\ }\bibfield  {title} {\bibinfo {title} {Photocarrier generation from interlayer charge-transfer transitions in {{WS2-graphene}} heterostructures},\ }\href {https://doi.org/10.1126/sciadv.1700324} {\bibfield  {journal} {\bibinfo  {journal} {Science Advances}\ }\textbf {\bibinfo {volume} {4}},\ \bibinfo {pages} {e1700324} (\bibinfo {year} {2018})}\BibitemShut {NoStop}%
\bibitem [{\citenamefont {Giusca}\ \emph {et~al.}(2019)\citenamefont {Giusca}, \citenamefont {Lin}, \citenamefont {Terrones}, \citenamefont {Gaskill}, \citenamefont {{Myers-Ward}},\ and\ \citenamefont {Kazakova}}]{giusca_probing_2019}%
  \BibitemOpen
  \bibfield  {author} {\bibinfo {author} {\bibfnamefont {C.~E.}\ \bibnamefont {Giusca}}, \bibinfo {author} {\bibfnamefont {Z.}~\bibnamefont {Lin}}, \bibinfo {author} {\bibfnamefont {M.}~\bibnamefont {Terrones}}, \bibinfo {author} {\bibfnamefont {D.~K.}\ \bibnamefont {Gaskill}}, \bibinfo {author} {\bibfnamefont {R.~L.}\ \bibnamefont {{Myers-Ward}}},\ and\ \bibinfo {author} {\bibfnamefont {O.}~\bibnamefont {Kazakova}},\ }\bibfield  {title} {\bibinfo {title} {Probing exciton species in atomically thin {{WS2}}{\textendash}graphene heterostructures},\ }\href {https://doi.org/10.1088/2515-7639/aafddb} {\bibfield  {journal} {\bibinfo  {journal} {Journal of Physics: Materials}\ }\textbf {\bibinfo {volume} {2}},\ \bibinfo {pages} {025001} (\bibinfo {year} {2019})}\BibitemShut {NoStop}%
\bibitem [{\citenamefont {Coy~Diaz}\ \emph {et~al.}(2015)\citenamefont {Coy~Diaz}, \citenamefont {Avila}, \citenamefont {Chen}, \citenamefont {Addou}, \citenamefont {Asensio},\ and\ \citenamefont {Batzill}}]{coy_diaz_direct_2015}%
  \BibitemOpen
  \bibfield  {author} {\bibinfo {author} {\bibfnamefont {H.}~\bibnamefont {Coy~Diaz}}, \bibinfo {author} {\bibfnamefont {J.}~\bibnamefont {Avila}}, \bibinfo {author} {\bibfnamefont {C.}~\bibnamefont {Chen}}, \bibinfo {author} {\bibfnamefont {R.}~\bibnamefont {Addou}}, \bibinfo {author} {\bibfnamefont {M.~C.}\ \bibnamefont {Asensio}},\ and\ \bibinfo {author} {\bibfnamefont {M.}~\bibnamefont {Batzill}},\ }\bibfield  {title} {\bibinfo {title} {Direct {{Observation}} of {{Interlayer Hybridization}} and {{Dirac Relativistic Carriers}} in {{Graphene}}/{{MoS2}} van der {{Waals Heterostructures}}},\ }\href {https://doi.org/10.1021/nl504167y} {\bibfield  {journal} {\bibinfo  {journal} {Nano Letters}\ }\textbf {\bibinfo {volume} {15}},\ \bibinfo {pages} {1135} (\bibinfo {year} {2015})}\BibitemShut {NoStop}%
\bibitem [{\citenamefont {Rohlfing}\ and\ \citenamefont {Louie}(2000)}]{rohlfing_electron-hole_2000}%
  \BibitemOpen
  \bibfield  {author} {\bibinfo {author} {\bibfnamefont {M.}~\bibnamefont {Rohlfing}}\ and\ \bibinfo {author} {\bibfnamefont {S.~G.}\ \bibnamefont {Louie}},\ }\bibfield  {title} {\bibinfo {title} {Electron-hole excitations and optical spectra from first principles},\ }\href {https://doi.org/10.1103/PhysRevB.62.4927} {\bibfield  {journal} {\bibinfo  {journal} {Physical Review B}\ }\textbf {\bibinfo {volume} {62}},\ \bibinfo {pages} {4927} (\bibinfo {year} {2000})}\BibitemShut {NoStop}%
\bibitem [{\citenamefont {R{\"o}sner}\ \emph {et~al.}(2016)\citenamefont {R{\"o}sner}, \citenamefont {Steinke}, \citenamefont {Lorke}, \citenamefont {Gies}, \citenamefont {Jahnke},\ and\ \citenamefont {Wehling}}]{rosner_two-dimensional_2016}%
  \BibitemOpen
  \bibfield  {author} {\bibinfo {author} {\bibfnamefont {M.}~\bibnamefont {R{\"o}sner}}, \bibinfo {author} {\bibfnamefont {C.}~\bibnamefont {Steinke}}, \bibinfo {author} {\bibfnamefont {M.}~\bibnamefont {Lorke}}, \bibinfo {author} {\bibfnamefont {C.}~\bibnamefont {Gies}}, \bibinfo {author} {\bibfnamefont {F.}~\bibnamefont {Jahnke}},\ and\ \bibinfo {author} {\bibfnamefont {T.~O.}\ \bibnamefont {Wehling}},\ }\bibfield  {title} {\bibinfo {title} {Two-{{Dimensional Heterojunctions}} from {{Nonlocal Manipulations}} of the {{Interactions}}},\ }\href {https://doi.org/10.1021/acs.nanolett.5b05009} {\bibfield  {journal} {\bibinfo  {journal} {Nano Letters}\ }\textbf {\bibinfo {volume} {16}},\ \bibinfo {pages} {2322} (\bibinfo {year} {2016})}\BibitemShut {NoStop}%
\bibitem [{\citenamefont {Stauber}\ \emph {et~al.}(2008)\citenamefont {Stauber}, \citenamefont {Peres},\ and\ \citenamefont {Geim}}]{stauber_optical_2008}%
  \BibitemOpen
  \bibfield  {author} {\bibinfo {author} {\bibfnamefont {T.}~\bibnamefont {Stauber}}, \bibinfo {author} {\bibfnamefont {N.~M.~R.}\ \bibnamefont {Peres}},\ and\ \bibinfo {author} {\bibfnamefont {A.~K.}\ \bibnamefont {Geim}},\ }\bibfield  {title} {\bibinfo {title} {Optical conductivity of graphene in the visible region of the spectrum},\ }\href {https://doi.org/10.1103/PhysRevB.78.085432} {\bibfield  {journal} {\bibinfo  {journal} {Physical Review B}\ }\textbf {\bibinfo {volume} {78}},\ \bibinfo {pages} {085432} (\bibinfo {year} {2008})}\BibitemShut {NoStop}%
\bibitem [{\citenamefont {Nair}\ \emph {et~al.}(2008)\citenamefont {Nair}, \citenamefont {Blake}, \citenamefont {Grigorenko}, \citenamefont {Novoselov}, \citenamefont {Booth}, \citenamefont {Stauber}, \citenamefont {Peres},\ and\ \citenamefont {Geim}}]{nair_fine_2008}%
  \BibitemOpen
  \bibfield  {author} {\bibinfo {author} {\bibfnamefont {R.~R.}\ \bibnamefont {Nair}}, \bibinfo {author} {\bibfnamefont {P.}~\bibnamefont {Blake}}, \bibinfo {author} {\bibfnamefont {A.~N.}\ \bibnamefont {Grigorenko}}, \bibinfo {author} {\bibfnamefont {K.~S.}\ \bibnamefont {Novoselov}}, \bibinfo {author} {\bibfnamefont {T.~J.}\ \bibnamefont {Booth}}, \bibinfo {author} {\bibfnamefont {T.}~\bibnamefont {Stauber}}, \bibinfo {author} {\bibfnamefont {N.~M.~R.}\ \bibnamefont {Peres}},\ and\ \bibinfo {author} {\bibfnamefont {A.~K.}\ \bibnamefont {Geim}},\ }\bibfield  {title} {\bibinfo {title} {Fine {{Structure Constant Defines Visual Transparency}} of {{Graphene}}},\ }\href {https://doi.org/10.1126/science.1156965} {\bibfield  {journal} {\bibinfo  {journal} {Science}\ }\textbf {\bibinfo {volume} {320}},\ \bibinfo {pages} {1308} (\bibinfo {year} {2008})}\BibitemShut {NoStop}%
\bibitem [{\citenamefont {Lorchat}\ \emph {et~al.}(2020)\citenamefont {Lorchat}, \citenamefont {L{\'o}pez}, \citenamefont {Robert}, \citenamefont {Lagarde}, \citenamefont {Froehlicher}, \citenamefont {Taniguchi}, \citenamefont {Watanabe}, \citenamefont {Marie},\ and\ \citenamefont {Berciaud}}]{lorchat_filtering_2020}%
  \BibitemOpen
  \bibfield  {author} {\bibinfo {author} {\bibfnamefont {E.}~\bibnamefont {Lorchat}}, \bibinfo {author} {\bibfnamefont {L.~E.~P.}\ \bibnamefont {L{\'o}pez}}, \bibinfo {author} {\bibfnamefont {C.}~\bibnamefont {Robert}}, \bibinfo {author} {\bibfnamefont {D.}~\bibnamefont {Lagarde}}, \bibinfo {author} {\bibfnamefont {G.}~\bibnamefont {Froehlicher}}, \bibinfo {author} {\bibfnamefont {T.}~\bibnamefont {Taniguchi}}, \bibinfo {author} {\bibfnamefont {K.}~\bibnamefont {Watanabe}}, \bibinfo {author} {\bibfnamefont {X.}~\bibnamefont {Marie}},\ and\ \bibinfo {author} {\bibfnamefont {S.}~\bibnamefont {Berciaud}},\ }\bibfield  {title} {\bibinfo {title} {Filtering the photoluminescence spectra of atomically thin semiconductors with graphene},\ }\href {https://doi.org/10.1038/s41565-020-0644-2} {\bibfield  {journal} {\bibinfo  {journal} {Nature Nanotechnology}\ }\textbf {\bibinfo {volume} {15}},\ \bibinfo {pages} {283} (\bibinfo {year} {2020})}\BibitemShut {NoStop}%
\bibitem [{\citenamefont {Deslippe}\ \emph {et~al.}(2012)\citenamefont {Deslippe}, \citenamefont {Samsonidze}, \citenamefont {Strubbe}, \citenamefont {Jain}, \citenamefont {Cohen},\ and\ \citenamefont {Louie}}]{deslippe_berkeleygw_2012}%
  \BibitemOpen
  \bibfield  {author} {\bibinfo {author} {\bibfnamefont {J.}~\bibnamefont {Deslippe}}, \bibinfo {author} {\bibfnamefont {G.}~\bibnamefont {Samsonidze}}, \bibinfo {author} {\bibfnamefont {D.~A.}\ \bibnamefont {Strubbe}}, \bibinfo {author} {\bibfnamefont {M.}~\bibnamefont {Jain}}, \bibinfo {author} {\bibfnamefont {M.~L.}\ \bibnamefont {Cohen}},\ and\ \bibinfo {author} {\bibfnamefont {S.~G.}\ \bibnamefont {Louie}},\ }\bibfield  {title} {\bibinfo {title} {{{BerkeleyGW}}: {{A}} massively parallel computer package for the calculation of the quasiparticle and optical properties of materials and nanostructures},\ }\href {https://doi.org/10.1016/j.cpc.2011.12.006} {\bibfield  {journal} {\bibinfo  {journal} {Computer Physics Communications}\ }\textbf {\bibinfo {volume} {183}},\ \bibinfo {pages} {1269} (\bibinfo {year} {2012})}\BibitemShut {NoStop}%
\bibitem [{\citenamefont {Fang}\ \emph {et~al.}(2022)\citenamefont {Fang}, \citenamefont {Yao}, \citenamefont {Zhang}, \citenamefont {Wang}, \citenamefont {Jiang}, \citenamefont {Huang}, \citenamefont {Korgel}, \citenamefont {Terrones}, \citenamefont {Al{\`u}},\ and\ \citenamefont {Zheng}}]{fang_room-temperature_2022}%
  \BibitemOpen
  \bibfield  {author} {\bibinfo {author} {\bibfnamefont {J.}~\bibnamefont {Fang}}, \bibinfo {author} {\bibfnamefont {K.}~\bibnamefont {Yao}}, \bibinfo {author} {\bibfnamefont {T.}~\bibnamefont {Zhang}}, \bibinfo {author} {\bibfnamefont {M.}~\bibnamefont {Wang}}, \bibinfo {author} {\bibfnamefont {T.}~\bibnamefont {Jiang}}, \bibinfo {author} {\bibfnamefont {S.}~\bibnamefont {Huang}}, \bibinfo {author} {\bibfnamefont {B.~A.}\ \bibnamefont {Korgel}}, \bibinfo {author} {\bibfnamefont {M.}~\bibnamefont {Terrones}}, \bibinfo {author} {\bibfnamefont {A.}~\bibnamefont {Al{\`u}}},\ and\ \bibinfo {author} {\bibfnamefont {Y.}~\bibnamefont {Zheng}},\ }\bibfield  {title} {\bibinfo {title} {Room-{{Temperature Observation}} of {{Near-Intrinsic Exciton Linewidth}} in {{Monolayer WS2}}},\ }\href {https://doi.org/10.1002/adma.202108721} {\bibfield  {journal} {\bibinfo  {journal} {Advanced Materials}\ }\textbf {\bibinfo {volume} {34}},\ \bibinfo {pages} {2108721} (\bibinfo {year} {2022})}\BibitemShut {NoStop}%
\bibitem [{\citenamefont {Selig}\ \emph {et~al.}(2016)\citenamefont {Selig}, \citenamefont {Bergh{\"a}user}, \citenamefont {Raja}, \citenamefont {Nagler}, \citenamefont {Sch{\"u}ller}, \citenamefont {Heinz}, \citenamefont {Korn}, \citenamefont {Chernikov}, \citenamefont {Malic},\ and\ \citenamefont {Knorr}}]{selig_excitonic_2016}%
  \BibitemOpen
  \bibfield  {author} {\bibinfo {author} {\bibfnamefont {M.}~\bibnamefont {Selig}}, \bibinfo {author} {\bibfnamefont {G.}~\bibnamefont {Bergh{\"a}user}}, \bibinfo {author} {\bibfnamefont {A.}~\bibnamefont {Raja}}, \bibinfo {author} {\bibfnamefont {P.}~\bibnamefont {Nagler}}, \bibinfo {author} {\bibfnamefont {C.}~\bibnamefont {Sch{\"u}ller}}, \bibinfo {author} {\bibfnamefont {T.~F.}\ \bibnamefont {Heinz}}, \bibinfo {author} {\bibfnamefont {T.}~\bibnamefont {Korn}}, \bibinfo {author} {\bibfnamefont {A.}~\bibnamefont {Chernikov}}, \bibinfo {author} {\bibfnamefont {E.}~\bibnamefont {Malic}},\ and\ \bibinfo {author} {\bibfnamefont {A.}~\bibnamefont {Knorr}},\ }\bibfield  {title} {\bibinfo {title} {Excitonic linewidth and coherence lifetime in monolayer transition metal dichalcogenides},\ }\href {https://doi.org/10.1038/ncomms13279} {\bibfield  {journal} {\bibinfo  {journal} {Nature Communications}\ }\textbf {\bibinfo {volume} {7}},\ \bibinfo {pages} {13279} (\bibinfo {year} {2016})}\BibitemShut {NoStop}%
\bibitem [{\citenamefont {Chan}\ \emph {et~al.}(2023)\citenamefont {Chan}, \citenamefont {Haber}, \citenamefont {Naik}, \citenamefont {Neaton}, \citenamefont {Qiu}, \citenamefont {{da Jornada}},\ and\ \citenamefont {Louie}}]{chan_exciton_2023}%
  \BibitemOpen
  \bibfield  {author} {\bibinfo {author} {\bibfnamefont {Y.-h.}\ \bibnamefont {Chan}}, \bibinfo {author} {\bibfnamefont {J.~B.}\ \bibnamefont {Haber}}, \bibinfo {author} {\bibfnamefont {M.~H.}\ \bibnamefont {Naik}}, \bibinfo {author} {\bibfnamefont {J.~B.}\ \bibnamefont {Neaton}}, \bibinfo {author} {\bibfnamefont {D.~Y.}\ \bibnamefont {Qiu}}, \bibinfo {author} {\bibfnamefont {F.~H.}\ \bibnamefont {{da Jornada}}},\ and\ \bibinfo {author} {\bibfnamefont {S.~G.}\ \bibnamefont {Louie}},\ }\bibfield  {title} {\bibinfo {title} {Exciton {{Lifetime}} and {{Optical Line Width Profile}} via {{Exciton}}{\textendash}{{Phonon Interactions}}: {{Theory}} and {{First-Principles Calculations}} for {{Monolayer MoS2}}},\ }\href {https://doi.org/10.1021/acs.nanolett.3c00732} {\bibfield  {journal} {\bibinfo  {journal} {Nano Letters}\ }\textbf {\bibinfo {volume} {23}},\ \bibinfo {pages} {3971} (\bibinfo {year} {2023})}\BibitemShut {NoStop}%
\bibitem [{\citenamefont {Olivero}\ and\ \citenamefont {Longbothum}(1977)}]{olivero_empirical_1977}%
  \BibitemOpen
  \bibfield  {author} {\bibinfo {author} {\bibfnamefont {J.~J.}\ \bibnamefont {Olivero}}\ and\ \bibinfo {author} {\bibfnamefont {R.~L.}\ \bibnamefont {Longbothum}},\ }\bibfield  {title} {\bibinfo {title} {Empirical fits to the {{Voigt}} line width: {{A}} brief review},\ }\href {https://doi.org/10.1016/0022-4073(77)90161-3} {\bibfield  {journal} {\bibinfo  {journal} {Journal of Quantitative Spectroscopy and Radiative Transfer}\ }\textbf {\bibinfo {volume} {17}},\ \bibinfo {pages} {233} (\bibinfo {year} {1977})}\BibitemShut {NoStop}%
\end{thebibliography}%
\end{document}